\documentclass[preprint2]{aastex}

\usepackage[hyperindex,breaklinks]{hyperref}
\usepackage{amssymb,amsmath,array}
\usepackage{multicol,threeparttable}
\usepackage{graphicx}
\graphicspath{{./figures/}}

\def\RG{{R_{\rm G}}}
\def\Porb{{P_{\rm orb}}}

\def\Hz{{\rm Hz}}

\def\c{{\rm c}}
\def\s{{\rm s}}

\def\figwd{.49\textwidth}
\newcommand{\figimg}[3][\figwd]{\parbox{#1}{\centering \ifx&#2&{}\else{#2.\\}\fi\includegraphics[width={#1}]{#3} }}

\usepackage{natbib}
\bibpunct{(}{)}{;}{}{}{,}

\slugcomment{}
\shorttitle{Variability from Nonaxisymmetric Fluctuations}
\shortauthors{Henisey et al.}

\begin{document}

\title{
Variability from Nonaxisymmetric Fluctuations Interacting with Standing Shocks in
Tilted Black Hole Accretion Disks
}

\author{Ken B. Henisey}
\affil{Natural Science Division, Pepperdine University, Malibu, CA 90263, USA}

\author{Omer M. Blaes}
\affil{Department of Physics, University of California, Santa Barbara, CA 93106, USA}

\and

\author{P. Chris Fragile}
\affil{Department of Physics and Astronomy, College of Charleston, Charleston, SC 29424, USA}

\begin{abstract}
We study the spatial and temporal behavior of fluid in fully three-dimensional, general 
relativistic, magnetohydrodynamical simulations of both tilted and untilted black hole accretion 
flows. We uncover characteristically greater variability in tilted simulations at frequencies 
similar to those predicted by the formalism of trapped modes, but ultimately conclude that its 
spatial structure is inconsistent with a modal interpretation. We find instead that previously 
identified, transient, over-dense clumps orbiting on roughly Keplerian trajectories appear 
generically in our global simulations, independent of tilt. Associated with these fluctuations are 
acoustic spiral waves interior to the orbits of the clumps. We show that the two nonaxisymmetric 
standing shock structures that exist in the inner regions of these tilted flows effectively amplify
the variability caused by these spiral waves to markedly higher levels than in untilted flows, which 
lack standing shocks. Our identification of clumps, spirals, and spiral-shock interactions in these 
fully general relativistic, magnetohydrodynamical simulations suggests that these features may be
important dynamical elements in models which incorporate tilt as a way to explain the observed
variability in black hole accretion flows.
\end{abstract}


\section{Introduction}

Much current theoretical modeling of black hole accretion flows is done in the context of the 
magnetorotational instability (MRI; \citealt{bal91}) which drives MHD turbulence and facilitates the 
transport of angular momentum. However, there are few, if any, observationally testable 
manifestations of MRI turbulence. Because turbulence is inherently time-dependent, one would hope 
that variability would be a promising avenue to explore, especially given that it is ubiquitous in 
observations across the electromagnetic spectrum of accreting black hole sources. Particularly in 
the X-rays, observed variability power spectra consist of continua that can be fit by broken power 
laws, as well as discrete quasi-periodic oscillations (QPOs). In black hole X-ray binaries, both 
the amplitude and slopes of the continua, as well as the presence and character of the QPOs, depend 
on the observed X-ray spectral state (e.g. \citealt{rem06}). Low-frequency QPOs are observed in 
both the low/hard and the steep power law states, while high-frequency QPOs are only observed in 
the steep power law state. The high/soft or thermal state exhibits much less X-ray variability 
overall, and what variability it does exhibit appears to be associated with the hard X-ray, 
nonthermal component \citep{chu01}. 

Global simulations of the accretion flow with fully developed MRI turbulence have been used to model 
the expected continuum variability under a variety of assumptions that relate variations in local 
fluid quantities to variations of what might be observed (e.g. \citealt{nob09} and references 
therein). QPOs have also been looked for in such simulations, with mixed results. For example, 
\citet{rey09} failed to find the trapped inertial waves predicted by hydrodynamic disk models (e.g. 
\citealt{wag99,kat01}) in their global MHD simulations of accretion disks. However, similar 
simulations by \citet{one11} identified radially extended, low-frequency quasi-periodic behavior in 
the azimuthal magnetic field associated with dynamo cycles within the MRI turbulence itself, 
although how that would manifest itself observationally is unclear. Hydrodynamically unstable inner 
tori that re-establish themselves after magnetically mediated accretion episodes have been observed 
in MHD simulations \citep{mac08}, and may be a mechanism for low-frequency QPOs. High-frequency 
QPOs in the mass fluxes have also been claimed in at least one MHD simulation \citep{katy04}. 
\citet{sch06} applied various post-processing emission models and ray tracing calculations to a 
global MHD simulation to make predictions of observed variability power spectra, and found that 
transient QPOs could be seen, but they were likely insignificant as they would only appear in 
certain observer directions at any given time. On the other hand, \citet{dol12} recently discovered 
transient QPOs in their radiative post-processing of a global MHD simulation appropriate for the 
Galactic center source Sgr~A$^*$. These QPOs appear to originate from $m=1$ non-axisymmetric 
structures formed within the inner accretion flow. 

All the simulations used in these studies represent flows in which the angular momentum of the 
accretion flow is aligned with the spin of the black hole, or the black hole is not spinning at all. 
Tilted accretion flows, in which the angular momenta of the accretion flow and the spinning black 
hole are not aligned, have additional complexity in their dynamics \citep{fra07,fra08}, and 
therefore also likely in their variability. Orbits of fluid elements in tilted flows are generally 
eccentric, not circular, and the radial gradient of eccentricity can give rise to a pair of 
non-axisymmetric standing shocks which completely alter the character of the innermost parts of the 
flow. Provided the flow is not too geometrically thin, Lense-Thirring torques can cause global, 
rigid-body precession of the otherwise differentially rotating and turbulent flow. Tilted accretion 
flows are probably common around supermassive black holes in galactic nuclei, where the fuel source 
is unlikely to know about the spin of the black hole, and in X-ray binaries if the spin of the black 
hole is misaligned with respect to the binary orbital angular momentum. 

Motivated in part by predictions that hydrodynamic inertial modes can be excited by eccentric orbits 
and warps in a tilted disk \citep{kats04,kat08,fer08}, we examined variability power spectra of the
local fluid density, radial velocity, and vertical velocity in a simulation of a flow with an 
initial tilt angle of $15^\circ$ around a Kerr black hole with a dimensionless spin parameter 
$a/M=0.9$, as well as an untilted flow around the same black hole \citep{hen09}. In the tilted 
simulation, we identified a particularly strong, coherent feature in all three fluid variables that 
had the same frequency over an extended range of radii. The frequency of this feature was consistent 
with it being a trapped inertial mode, but we were unable to convincingly demonstrate that this was 
its true nature. It appeared to be associated with an overdense clump orbiting with the background 
flow, as well as inertial and inertial-acoustic waves. We therefore suggested that this was a 
preliminary confirmation that disk warps and eccentricity might excite inertial modes even in the 
presence of MRI turbulence. \citet{dex11} later did a radiative post-processing analysis of this 
tilted simulation and failed to find any manifestation of this discrete frequency feature in the 
simulated light curves that might be observed at infinity. 

Despite our failure in finding a radiative signal of this QPO, there is still much to be gleaned
from these simulations. The variability dynamics of turbulent accretion flows, and in particular
tilted accretion flows, remains very poorly understood. To that end, this paper extends our previous
analysis to a series of simulations, with longer durations and with 
$0^\circ$, $10^\circ$ and $15^\circ$ tilt angles. The tilted accretion flows generally exhibit 
discrete frequency features that have broad radial extents. However, these features are transient,
and the frequencies that are present come and go as the simulations progress. The features do not 
appear to be manifestations of trapped inertial waves, but are instead due to transient co-orbiting 
clumps that have associated radially extended waves. Such clumps are also present in the untilted 
simulations, but they are not accompanied by the same high-power radially extended structures. The
radial power arises in the tilted geometries owing to the standing shocks in the inner parts of the
flow which amplify the variability of each clump's associated wave. This amplification occurs
because the positive density fluctuations in the waves are significantly enhanced at the points
where they cross the shocks.

We present the basis for these conclusions as follows. In Section 2, we provide an overview of the 
simulations used in our analysis. In Section~\ref{sec:bg}, we describe the time-averaged structure 
of the inner flow regions, focusing in particular on the standing shocks in the tilted flows. In 
Section~\ref{sec:fp}, we extend our previous study \citep{hen09} and comment on the radial 
distribution of variability in our datasets. We present our analysis of the physics behind the 
variability in Section~\ref{sec:structure}, which consists of two parts. In Section~\ref{subsec:mode3d}, we present 
visualizations of the three-dimensional structure of the variability at a given frequency in the 
context of the nonaxisymmetric background. Motivated by the dominant $m=1$ modes originally 
identified in our previous work, we break down the variability into prograde and retrograde $m=1$ 
structures in Section~\ref{subsec:spirals} and discuss a possible mechanism by which these 
structures form. We summarize our conclusions in Section~\ref{sec:conclusions}. 

\section{Simulations}

This work aims to characterize variability in four distinct general relativistic, 
magnetohydrodynamic simulation datasets of black hole accretion flows generated by the Cosmos++ code 
\citep{ann05}. We continue our previous study \citep{hen09} of two such simulations originally 
introduced in \cite{fra07}: 90h, an untilted configuration with an initial fluid angular momentum 
aligned with the spin axis of an $a/M=0.9$ Kerr black hole, and 915h, a tilted flow with initial 
fluid angular momentum misaligned from the spin axis of the hole by $15^\circ$. We also study two 
new high-resolution simulations, 910h and 915h-64, both exploring tilted geometries with disk-spin 
misalignments of $10^\circ$ and $15^\circ$, respectively. These simulations all run on essentially 
the same grids in which the simulation domains reach equivalent peak resolutions of $128^3$ zones 
\citep{fra07}.

Each black hole spacetime is modeled using a Kerr-Schild coordinate system with spin-parameter 
$a/M=0.9$ and which is tilted to achieve the desired angle with respect to the grid midplane. We 
embed each spacetime with an analytic solution for a torus whose inner edge and pressure maximum lie 
in the grid midplane at $15\RG$ and $25\RG$, respectively, where $\RG\equiv GM/c^2$, and has an 
initial specific angular momentum distribution given by a power-law with exponent $q=1.68$ 
\citep{cha85}. Each torus was then seeded with different initial random perturbations. For this 
reason, simulations 915h and 915h-64 are fully independent, despite having identical misalignments 
and unperturbed initial tori. Weak poloidal magnetic field loops having $\beta_{\rm mag}\equiv 
P/P_{\rm mag}\ge 10$ lie along the isobars within the torus and ultimately seed the MRI. Integration 
from these initial conditions utilizes Cosmos++'s internal-energy, artificial viscosity formulation 
while the magnetic fields evolve by an advection-split form, using a hyperbolic divergence cleanser 
to maintain a roughly divergence-free field. For more detail, see \cite{fra05} and \cite{fra07}. 

Throughout this paper, an ``orbit'' refers to the orbital period of a circular geodesic in the 
equatorial plane of the black hole at the radius $25\RG$ of the pressure maximum of the initial 
torus: $1\Porb\simeq791 GM/c^3$. We begin our timing analysis four orbits after the start of each 
simulation, so that MRI turbulence is fully developed in the inner regions of the flow and the 
average mass accretion rate as a function of radius reaches a more or less uniform profile inside of 
$15\RG$. Datasets from simulations 90h, 910h, and 915h then include information from the next six 
orbits, that is, from time $t=4$ to $10\Porb$ of the evolution. The 915h-64 dataset, on the other 
hand, includes $12$ orbits of data split into two epochs labeled 915h-64a and 915h-64b, encompassing 
$t=4$ to $10\Porb$ and $t=10$ to $16\Porb$, respectively. A summary of the differences between these 
five datasets appears in Table~\ref{tab:sim_sum}.

\begin{table}
  \centering
  \begin{threeparttable}
    \caption[Summary of datasets and their respective tilts and epochs.]{Dataset Summary}
    \begin{tabular*}{\columnwidth}{@{\extracolsep{\fill}}cccr@{\extracolsep{0pt}$\;\le t <\;$}l}
      \hline\hline
      Dataset & Simulation & Tilt & \multicolumn{2}{c}{Time Domain\tnote{a}} \\
      \hline
      90h  & 90h &  $0^\circ$ & $4$ & $10$ \\
      910h & 910h & $10^\circ$ & $4$ & $10$ \\
      915h & 915h & $15^\circ$ & $4$ & $10$ \\
      915h-64a\tnote{b} & 915h-64 & $15^\circ$ & $4$ & $10$ \\
      915h-64b\tnote{b} & 915h-64 & $15^\circ$ & $10$ & $16$ \\
      \hline
    \end{tabular*}
    \begin{tablenotes}
      \item[a]{In units of the orbital period of a test particle at $25\RG$.}
      \item[b]{Though computationally continuous, we divide the analysis of simulation
       915h-64 into two, $6$ orbit epochs.}
    \end{tablenotes}
    \label{tab:sim_sum}
  \end{threeparttable}
\end{table}

\section{Background Flow Structure}\label{sec:bg}

A tilted accretion flow around a spinning black hole has no axis of symmetry, and the spatial 
structure of the flow can be quite complex. \cite{fra07} identified two high density arms that they 
called "plunging streams" in tilted 915h simulation originating at around $4\RG$ and dominating the 
accretion in the innermost regions of the flow. \cite{fra08} showed that two standing shocks 
accompany these overdense arms and extend outward to larger radii. They also demonstrated a sharp 
decrease in angular momentum inside $6\RG$ and indicated enhanced energy dissipation inside roughly 
$10\RG$ in comparison with the untilted flow, phenomena that they suggested are directly related to 
the shock structures.

As we shall see, these shock structures play an important role in the variability of tilted 
accretion flows. We therefore spend some time here discussing their structure in some detail for 
each of the tilted simulation datasets, using the rate of change of entropy as a diagnostic. For an 
ideal gas up to an arbitrary constant of integration, we may define the entropy such that 
$s\equiv\log{(p/\rho^\gamma)}$, where $\gamma$ is the fluid's adiabatic index, and all quantities 
are measured in the fluid rest frame. In our simulations, $\gamma=5/3$. Since the spatial location 
of a fluid element with rapidly increasing entropy strongly suggests the location of a shock front, 
we locate spatial maxima in the average rate of change of specific entropy calculated at each grid 
zone, 
\begin{align}
\frac{\Delta s}{\Delta t}&\equiv\left\langle\frac{u^\mu\nabla_\mu s}{u^t}\right\rangle_t,
\label{eq:shocks}
\end{align}
where $u^\mu$ is the fluid 4-velocity, $u^t$ is the time component of that 4-velocity, and 
$\nabla_\mu$ is a covariant derivative, which in the case of the scalar operand $s$, reduces to 
partial derivatives in the four spacetime coordinates. Here and throughout this paper, the angle 
brackets indicate averages over the subscripting spacetime coordinates and imply that the resulting 
expression remains a function of the unaveraged coordinates. Equation (\ref{eq:shocks}) is 
equivalent to $\Delta s/\Delta t=\langle\partial s/\partial t+\vec{v}\cdot\nabla s\rangle_t$, where 
$\vec{v}$ is the fluid's coordinate 3-velocity (that is, $v^i=u^i/u^t$) and $\nabla$ is a 3-gradient 
in the spatial coordinates. In this form, it is more clear that this represents the average 
Lagrangian rate of change of the entropy with respect to coordinate time, that is, the total change 
in entropy of fluid elements as they pass through a given grid zone over the total simulation time 
$[\int(ds/dt)dt]/\int dt$ or simply $\Delta s/\Delta t$. Since the accretion flow experiences 
roughly $24^\circ$ of Lense-Thirring precession during the time evolution in each of our datasets, 
features identified by time-averaged quantities like those defined above may be smeared by at most 
that angle.

\begin{figure*}
\centering
\def\figwd{.45\textwidth}
\begin{tabular*}{\textwidth}{@{\extracolsep{\fill}}*{2}{@{}b{\figwd}}@{}}
\figimg[\figwd]{a}{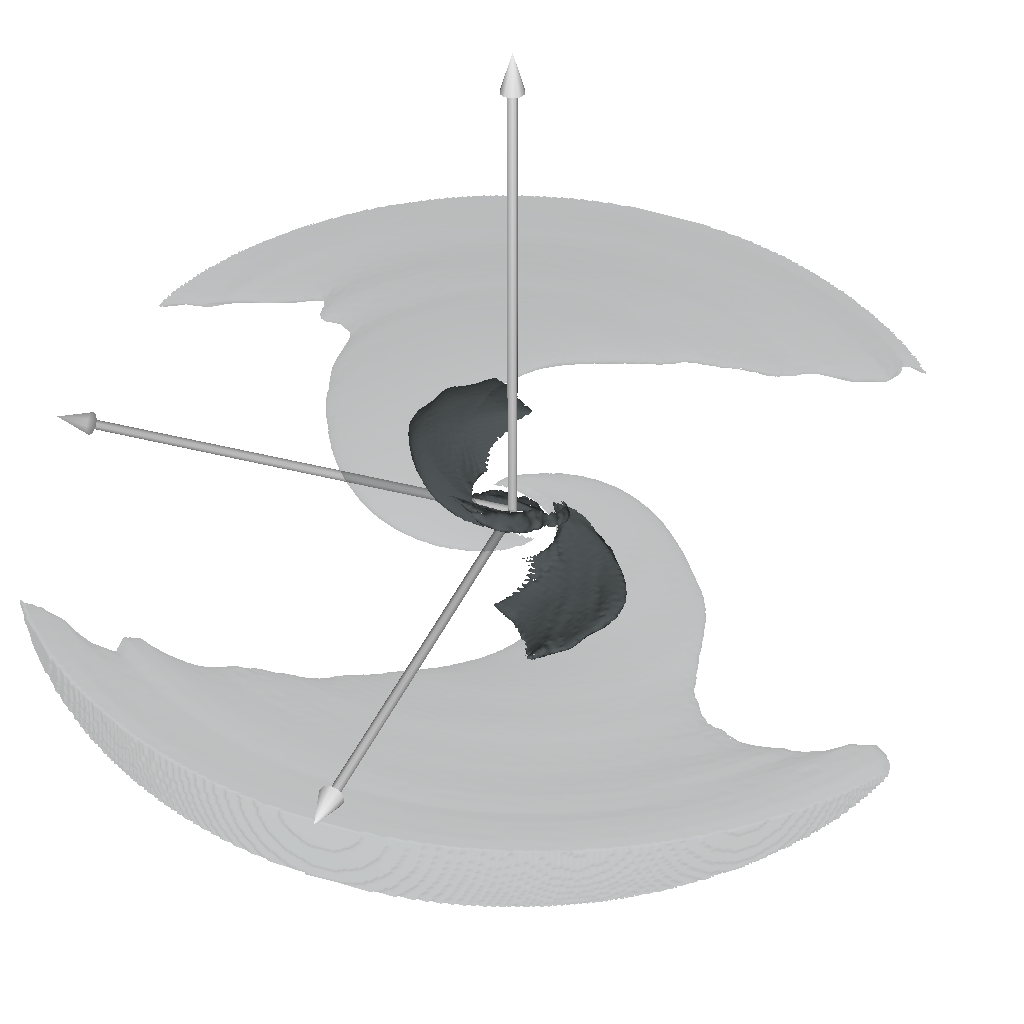} &
\figimg[\figwd]{b}{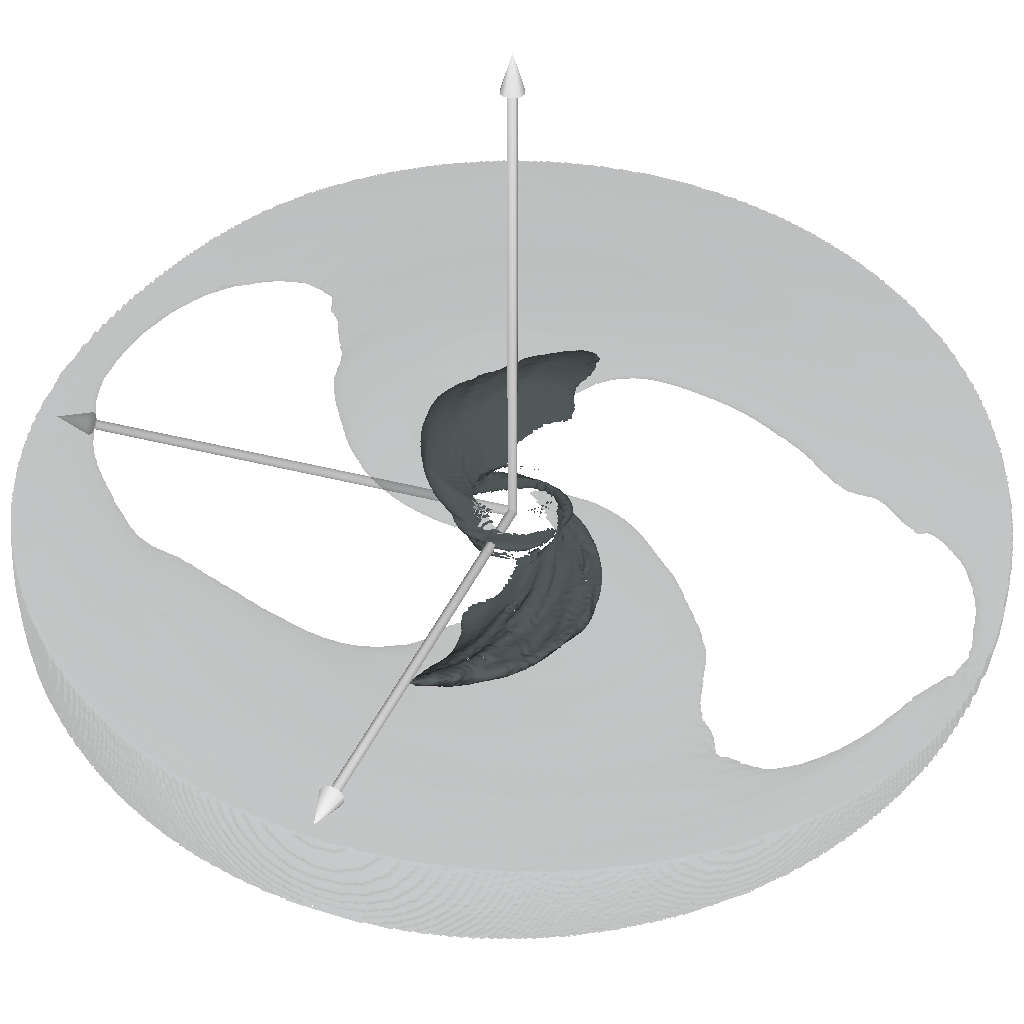} \\
\figimg[\figwd]{c}{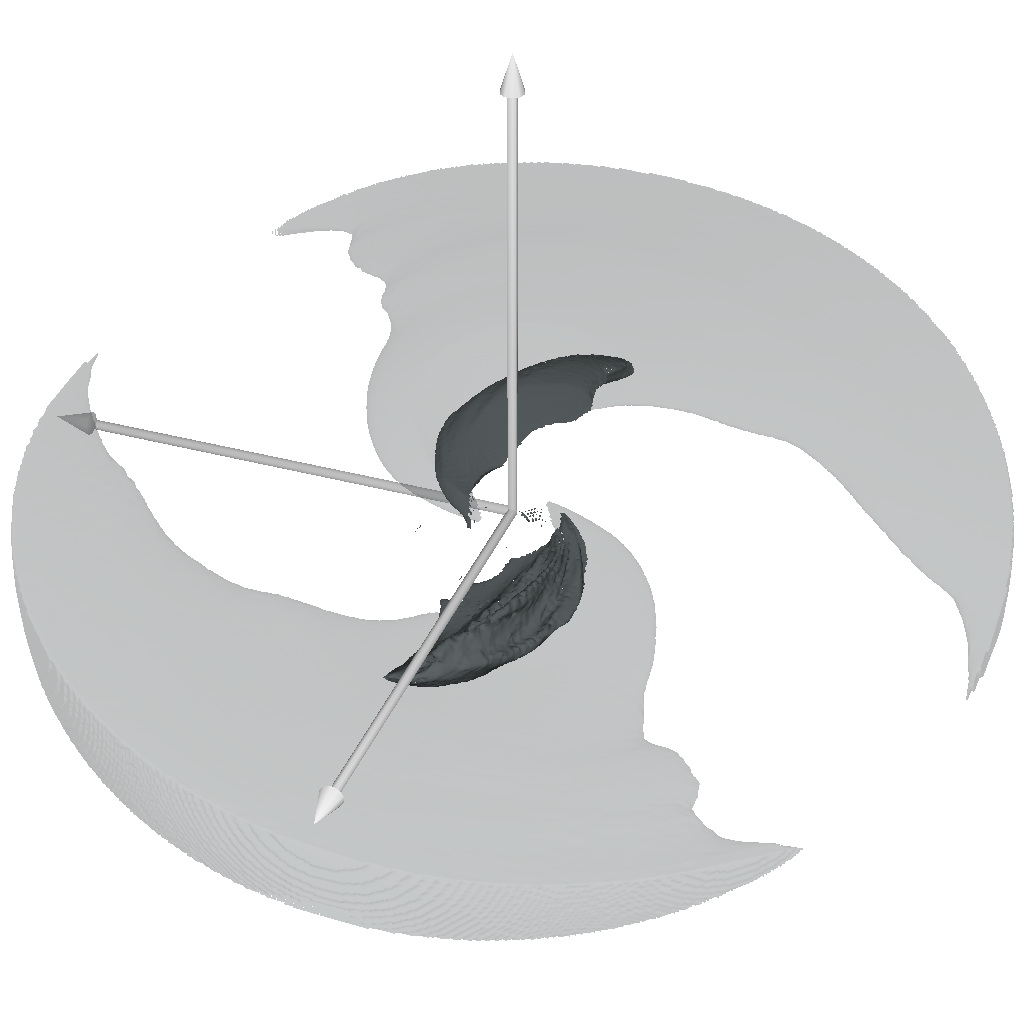} &
\figimg[\figwd]{d}{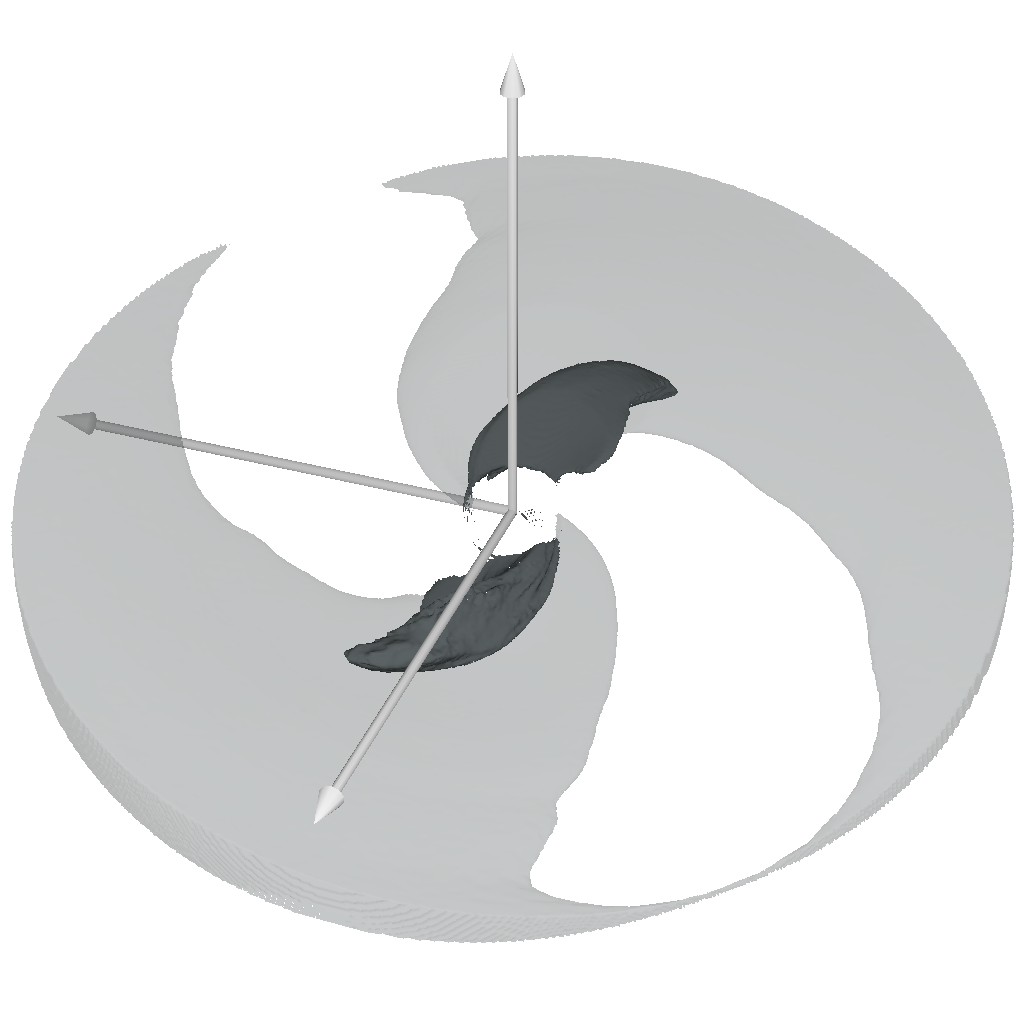}
\end{tabular*}
\caption[Three-dimensional density and shock structures in tilted simulations.]
 {Three-dimensional contours in time-averaged density normalized to the time and shell-averaged density
  $\langle\rho\rangle_t/\langle\rho\rangle_{t,\theta,\phi}$ (semi-transparent gray) and Lagrangian
  specific entropy generation rate $\Delta s/\Delta t$ in arbitrary units [solid gray; see
  Equation~\ref{eq:shocks}] drawn at $2.5$ and $0.07$, respectively, from 910h (a), 915h (b),
  915h-64a (c), and 915h-64b (d). Axes extend to $15\RG$. The regions enclosed in the
  entropy generation rate contour identify the location of standing shock surfaces within the
  flow.
  \label{fig:shocks}}
\end{figure*}

In Figure~\ref{fig:shocks}, we plot two contours: first, a semi-transparent contour inside of which 
$\langle\rho\rangle_t/\langle\rho\rangle_{t,\theta,\phi}>2.5$, that is, where the time-averaged 
density at a given point in space is $2.5$ times larger than the shell-averaged density at the same 
radius; second, a solid contour inside of which the time-averaged specific entropy generation rate 
$\Delta s/\Delta t>0.070$ in the same arbitrary units defined above. For comparison, 
Table~\ref{tab:shocks} details the average generation rates per coordinate volume, again in the same
arbitrary units, both inside the contours and outside (but still within $15\RG$). Our contour choice 
clearly contains dramatically larger specific entropy generation rates than the rest of the 
simulation volumes and therefore effectively locates these quite planar shock surfaces. Note that 
the sense of fluid orbital motion is right-handed about the vertical axes shown in 
Figure~\ref{fig:shocks}, and that at a given radius, the density is maximal just behind the standing 
shocks. 

\begin{table}
  \centering
  \begin{threeparttable}
    \caption[Specific entropy generation rates within standing shocks and characteristic radii.]
      {Specific entropy generation rate $\Delta s/\Delta t$ per coordinate volume}
    \begin{tabular*}{\columnwidth}{@{\extracolsep{\fill}}lcc}
      \hline\hline
      Dataset & Outside\tnote{a} & Inside\tnote{a} \\
      \hline
      910h     & 0.0011 & 0.123 \\
      915h     & 0.0027 & 0.132 \\ 
      915h-64a & 0.0024 & 0.138 \\
      915h-64b & 0.0017 & 0.154 \\
      \hline
    \end{tabular*}
    \begin{tablenotes}
      \item[a]{The average generation rate per coordinate volume inside and outside of the
               contours $\Delta s/\Delta t=0.07$ in Figure~\ref{fig:shocks} and with $r<15\RG$.}
    \end{tablenotes}
    \label{tab:shocks}
  \end{threeparttable}
\end{table}

\begin{figure*}
\centering
\begin{tabular*}{\textwidth}{@{\extracolsep{\fill}}*{2}{@{}b{\figwd}}@{}}
\multicolumn{2}{c}{\figimg{a}{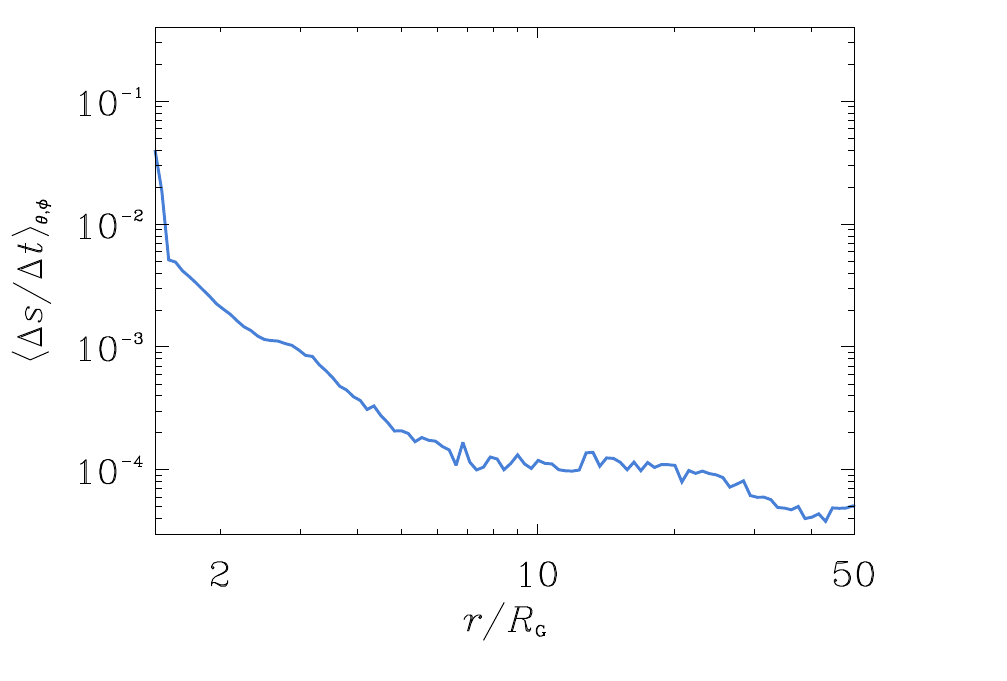}} \\
\figimg{b}{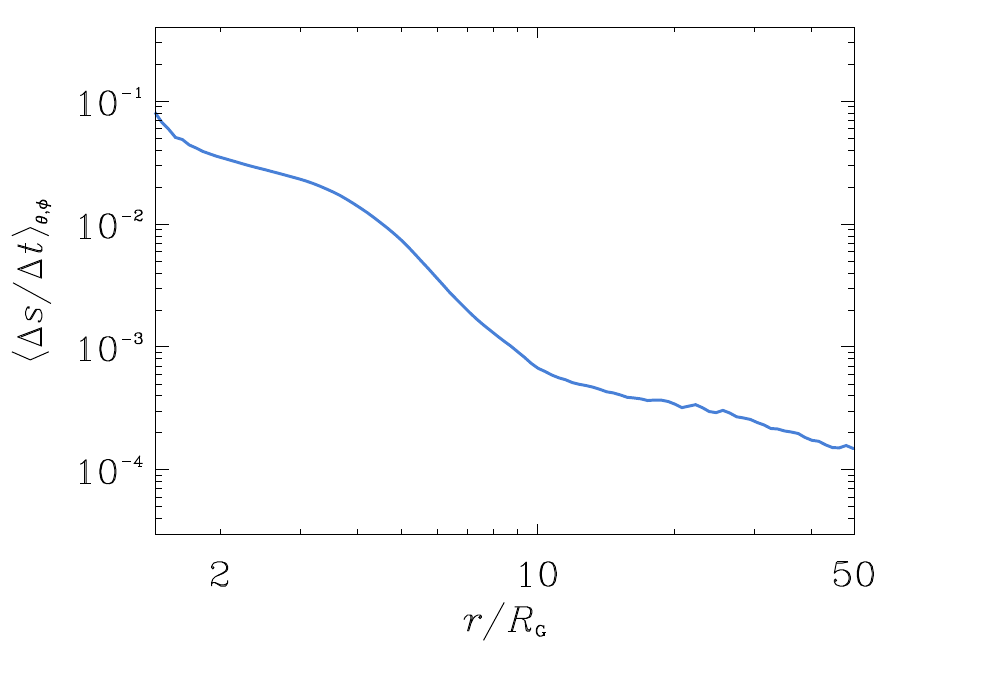} &
\figimg{c}{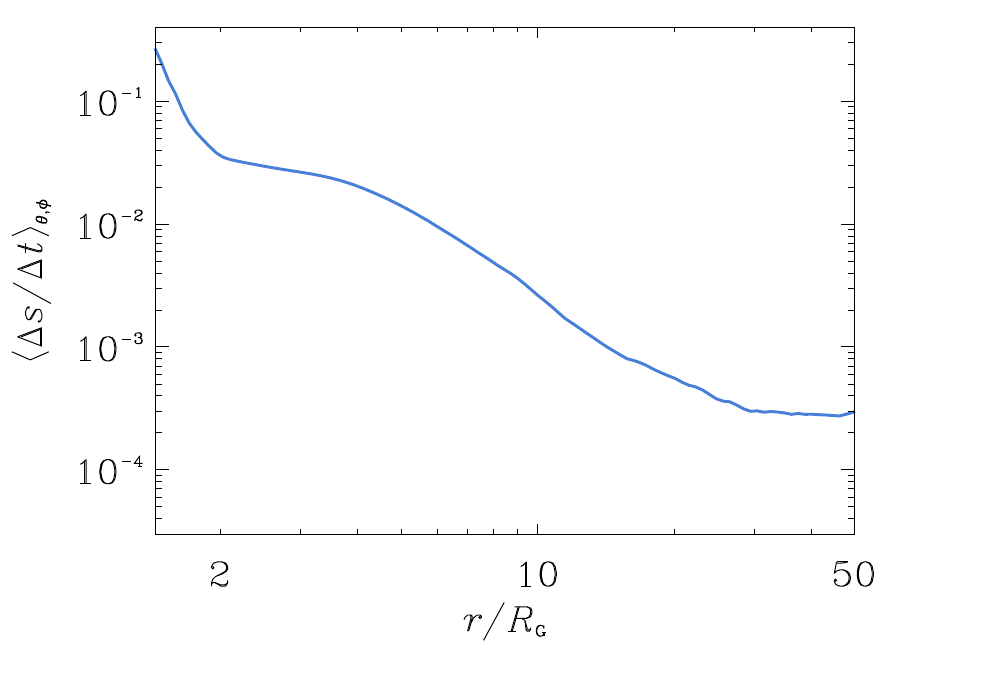} \\
\figimg{d}{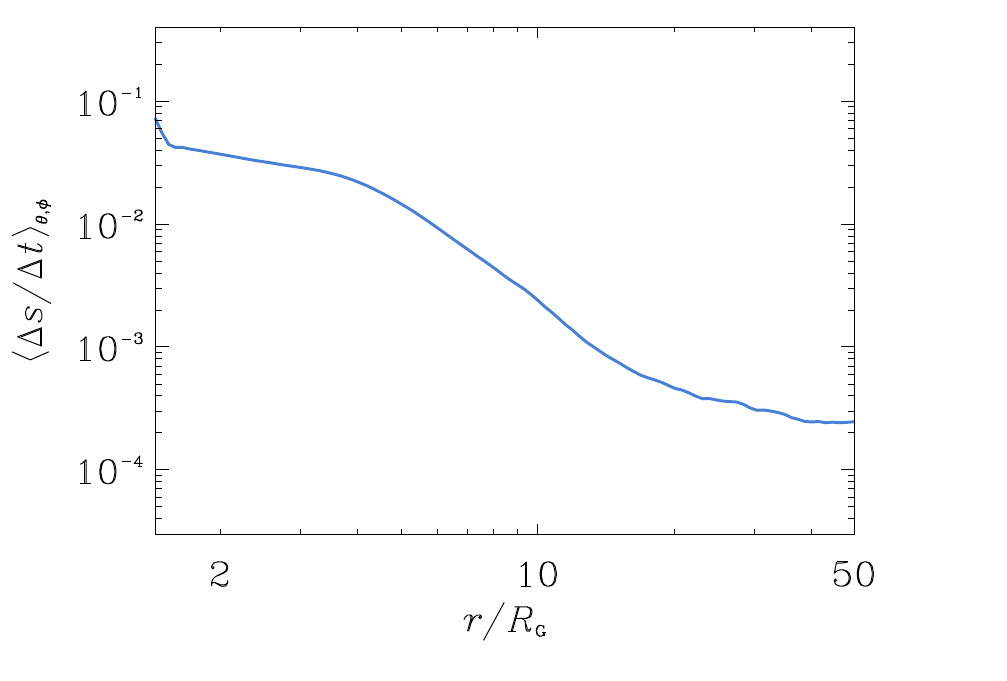} &
\figimg{e}{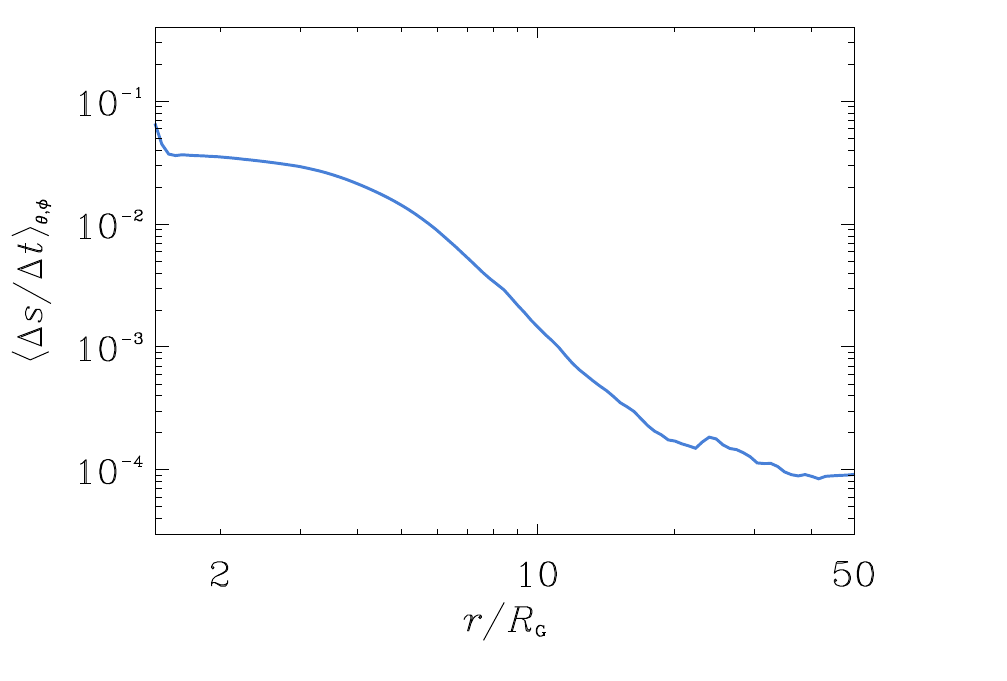}
\end{tabular*}
\caption[Time and shell-averaged specific entropy generation rate.]
 {Time and shell-averaged specific entropy generation rate as a function of radius in 90h (a),
  910h (b), 915h (c), 915h-64a (d), and 915h-64b (e).
  \label{fig:shock_str}}
\end{figure*}

We characterize the extent and strengths of the shocks by studying the shell and time-averaged 
Lagrangian specific entropy generation rates, shown in Figure~\ref{fig:shock_str}. For completeness, 
this figure also includes a panel for the untilted simulation which exhibits generation rates at 
least $50$ times smaller than those in the tilted simulations; this is consistent with the absence 
of standing shocks in that dataset. Each tilted simulation can be characterized by a transition from 
an almost constant plateau at smaller radii to the almost constant baseline at larger radii. This 
transition occurs more sharply in the less tilted 910h dataset. Additionally, the center of that 
transition region occurs at a smaller radius, around $7\RG$, than in 915h and both epochs of 915h-64 
where it takes place at around $10\RG$. As a result, the total time and volume integrated specific 
entropy generation rate inside the transition region is smallest in 910h and largest in 915h-64b. 
Since orbital time scales away from the hole are not necessarily small compared to the simulations' 
durations, one can see subtle differences at large radii in the entropy generation rate between 
915h-64a and 915h-54b (panels d and e, respectively) which indicate continuing disk evolution. 

In the $15^\circ$ tilted simulations, we measure compression ratios across these shock surfaces as 
high as $4.7$, depending on location. This is comparable to what we would expect: 
for a constant $5/3$ adiabatic index, as our simulations assume, hydrodynamic shocks should 
have a compression ratio of at most $4$, neglecting magnetic fields and relativity. We find
compression ratios as high as $4.0$ in the $10^\circ$ simulation \citep{gen12}.

\begin{figure*}
\centering
\def\figwd{.45\textwidth}
\begin{tabular*}{\textwidth}{@{\extracolsep{\fill}}*{2}{@{}b{\figwd}}@{}}
\figimg[\figwd]{a}{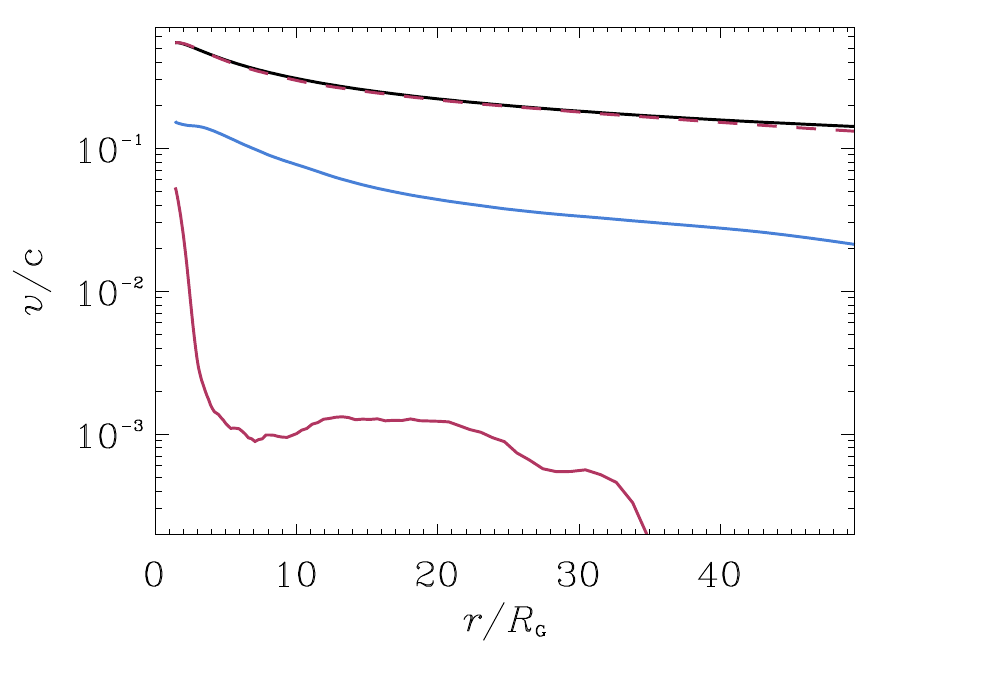} &
\figimg[\figwd]{b}{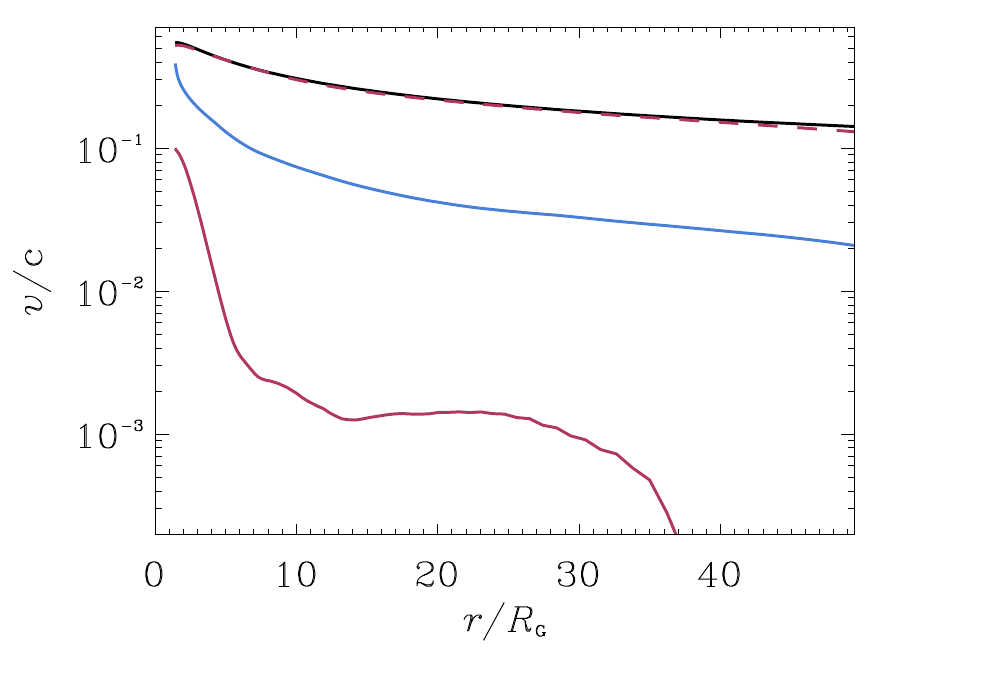}
\end{tabular*}
\caption[Radial behavior of characteristic velocities.]
 {The time and shell-averaged radial (solid red) and non-radial (dashed red;
  $r\sqrt{(v^\theta)^2+(v^\phi\sin\theta)^2}$) velocities, the time and shell-averaged sound speed
  (blue), and the test particle Keplerian orbital velocity (black) for simulations 90h (a) and 915h
  (b). Note that the non-radial component of the velocity very closely tracks the orbital velocity
  (dashed red and black, respectively) as expected. Time-averages span the interval $t=4$ to $10\Porb$ and shell
  averages include a weighting by the time-averaged density. In both simulations, we avoid
  characterizing the region between $4$ and $10\RG$ as ``plunging'' since fluid elements therein
  orbit the hole many times before making their final plunge toward the horizon.
  \label{fig:velocities}}
\end{figure*}

\cite{fra08} also presented comparative plots of characteristic velocities in 90h and 915h. 
Figure~\ref{fig:velocities} reproduces these plots, only now the time-averages cover the interval 
$t=4$ to $10\Porb$ instead of the smaller interval $t=7$ to $10\Porb$, and we have also added curves 
for the non-radial component of the velocity and the test particle orbital velocity. \citet{fra08} 
stated that the sharp upturn in $v^r$ indicated the start of the plunging region, and pointed out 
that this occurred at larger radii in the tilted simulation. Note, however, that fluid elements 
outside of roughly $4\RG$ in the 915h tilted simulation make many complete orbits before truly 
plunging toward the hole. The high density arms shown in Figure~\ref{fig:shocks} are therefore 
mostly due to postshock compression, and are not literally plunging streams, at least outside 
roughly $4\RG$. As we discuss below, we posit that this shock compression is at the heart of the 
enhanced variability from spiral waves associated with transient orbiting clumps. 

\section{Power Density Spectra}\label{sec:fp}

\begin{figure*}
\centering
\begin{tabular*}{\textwidth}{@{\extracolsep{\fill}}*{2}{@{}b{\figwd}}@{}}
\multicolumn{2}{c}{\centering \figimg{a}{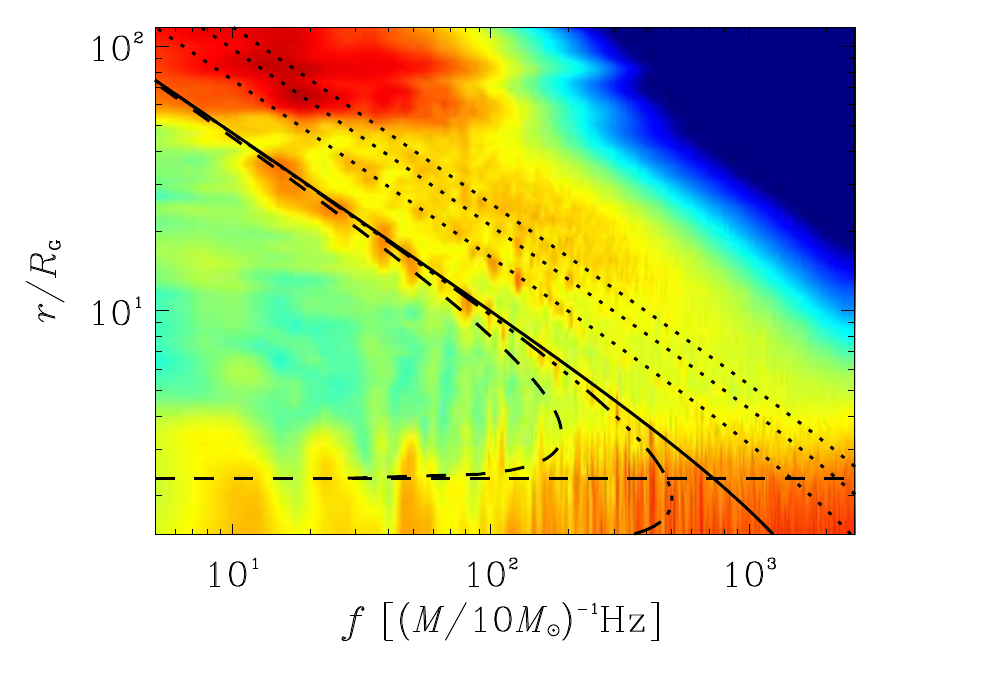}} \\
\figimg{b}{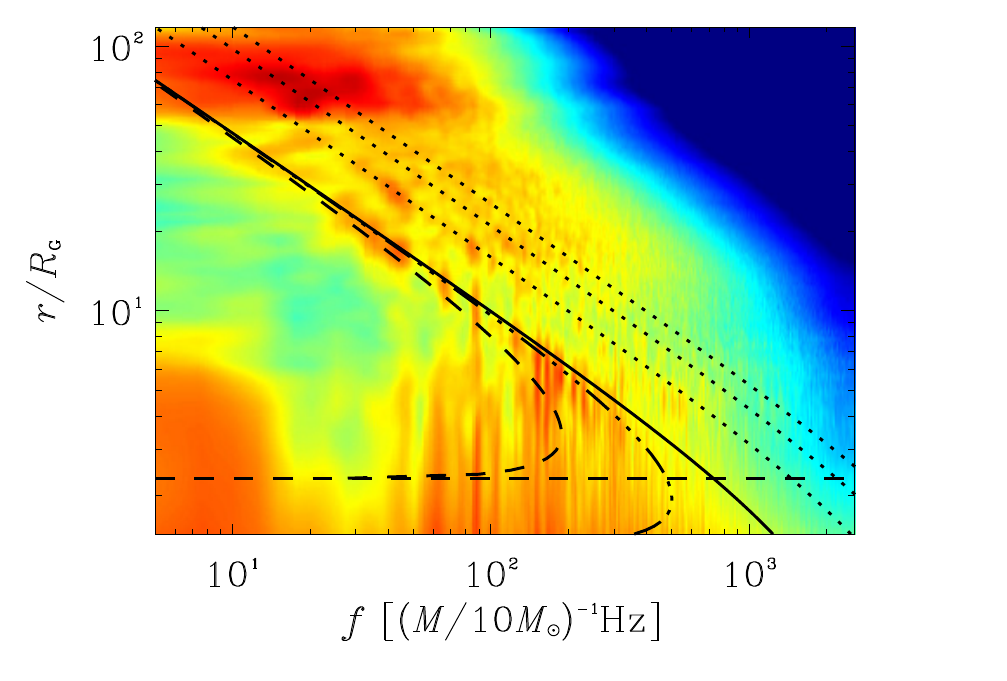} &
\figimg{c}{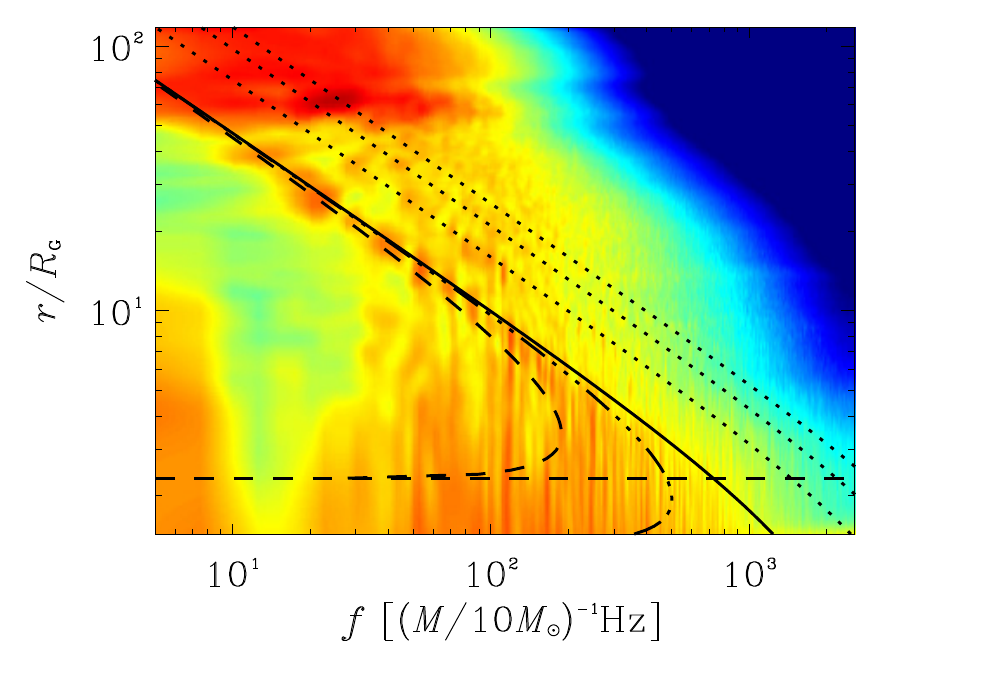} \\
\figimg{d}{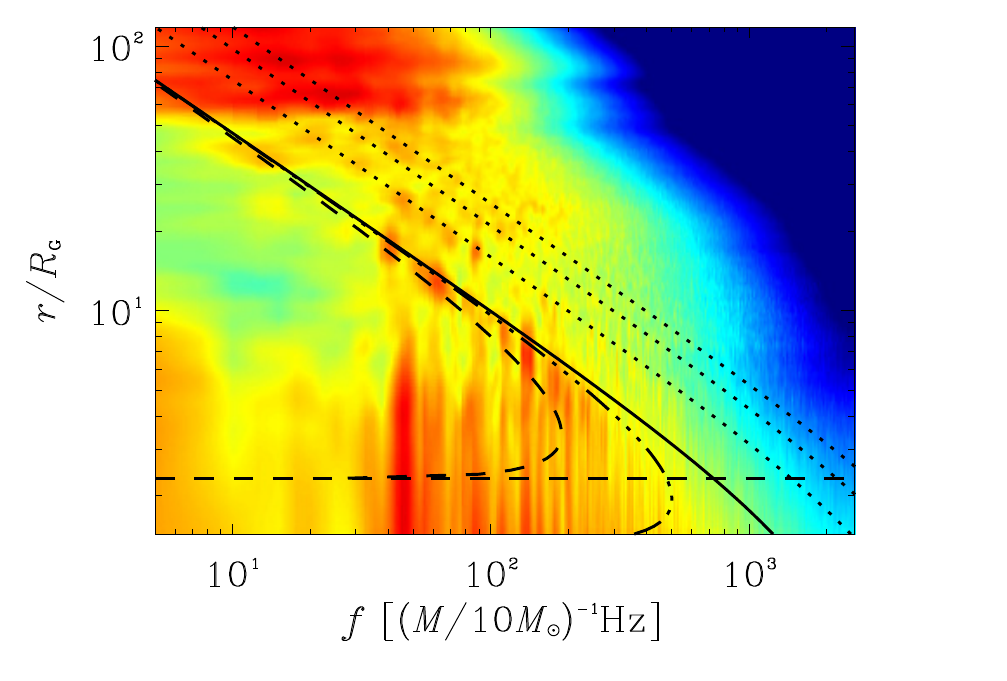} &
\figimg{e}{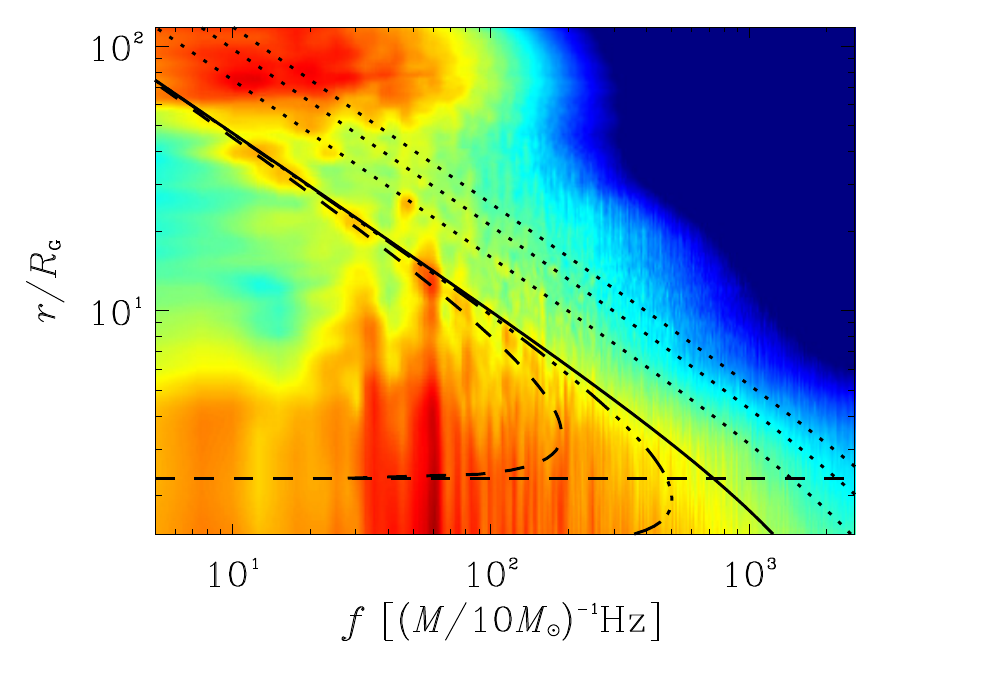} \\
\multicolumn{2}{c}{\centering \figimg{}{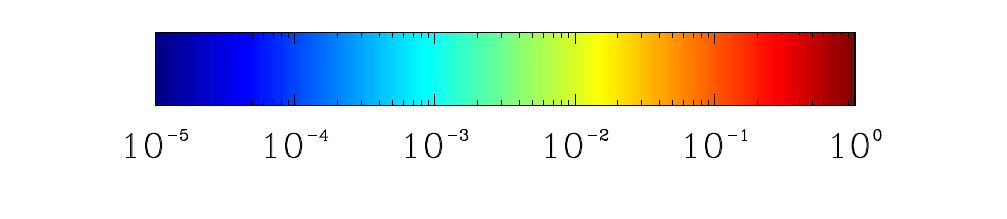}}
\end{tabular*}
\caption[Shell-averaged power spectra of density for all datasets.]
 {Shell-averaged power spectra of density $fP_\rho$ as a function of frequency $f$ and coordinate
  radius $r$ (cf.\ Equation~\ref{eq:pds_norm}) for datasets 90h (a), 910h (b), 915h (c),
  915h-64a (d), and 915h-64b (e). Over-plots include the orbital frequency (solid) and its
  harmonics (dotted), the geodesic radial epicyclic frequency and the ISCO radius (dashed), and the
  geodesic vertical epicyclic frequency (triple-dot dashed).
  \label{fig:fp_rho}}
\end{figure*}

Our first step in probing the three-dimensional structure of variability in our simulated accretion 
flows is to understand the radial distribution of spectral power. Interesting structures that
rotate, oscillate, or otherwise vary at characteristic frequencies may be identified in frequency
space as radially extended regions with enhanced power. To this end, we follow the method outlined
in our previous work \citep{hen09} and compute the shell-averaged spectral power in fluid density
fluctuations, 
\begin{eqnarray}
P_\rho(f,r)&\equiv\langle|\tilde{\rho}|^2\rangle_{\theta,\phi}/\langle\rho^2\rangle_{t,\theta,\phi},
\label{eq:pds_norm}
\end{eqnarray}
where $\tilde{\rho}$ is the temporal Fourier transform of the (prewhitened, windowed, and padded)
fluid density and is a function of space and frequency. We plot the resulting spectra for each of
the five datasets in Figure~\ref{fig:fp_rho}.

Two features in these spectra warrant discussion. First, in \cite{hen09} we showed that transient
clumps traveling on roughly Keplerian trajectories exist amid the MRI turbulence in simulations
90h and 915h. They appear as enhanced power along the curve of orbital frequency versus radius. In
Figure~\ref{fig:fp_rho}, we now clearly see that these orbiting clumps arise generically in all
five of our datasets.

Second, a set of features that we previously identified in 915h in \cite{hen09} likewise appear here 
strongly in each of the tilted simulations. Specifically, the tilted datasets exhibit significant 
power in vertical streaks extending inward in radius from the orbital frequency curve. For the 
remainder of this work, we will refer to such power as ``intraorbital'', denoting that it is power 
at a specific frequency and at all radii inward of the frequency's corotation radius. Importantly, 
these intraorbital, coherent structures coincide with the power enhancements along the orbital 
frequency curve associated with orbiting clumps. In the next section, we explore the relationship 
between these intraorbital features and their respective orbital clumps. 

In \cite{hen09}, we concentrated much of our study on the $118\Hz$ feature in the 915h simulation. 
Here we note that strong vertical tracks also appear in 910h at $150\Hz$, in 915h-64a at $140\Hz$, 
and in 915h-64b at $60\Hz$. Noting that 915h-64a and 915h-64b represent different epochs of the same
simulation and that 915h and 915h-64a differ only in there initial random perturbations, variety
in the position of and power in these simulations' vertical streaks indicates that while
these structures are long lived with respect to the local orbital period, they are nonetheless
transient features: they come and go on time scales smaller than but comparable to the simulation
durations. While these intraorbital streaks appear exclusively in the tilted simulations, features
that are similar in structure though markedly weaker (by perhaps as large a 
factor as 30) in spectral power exist within the untilted flow. One such example appears at 
$110\Hz$. Note that in all cases (both tilted and untilted), a local minimum seems to occur along 
the vertical track at roughly the midpoint between the corotation radius and innermost stable 
circular orbit (ISCO). 

These intraorbital power features contribute to generally enhanced variability in each of the tilted 
simulations at all frequencies greater than approximately $50\Hz$ (equivalently, the orbital 
frequency at roughly $15\RG$) and radially inward of the orbital frequency curve. There appears to 
be a qualitative trend of increasing intraorbital power with increasing tilt and with more advanced 
flow evolution, though we have not attempted to quantitatively assess its significance. 

To summarize, Figure~\ref{fig:fp_rho} confirms that transient over-dense clumps orbiting with the 
background flow are generic in our simulations. The mechanism responsible for the formation of these 
clumps remains unexplained. In terms of our global variability study however, it is sufficient that 
these clumps exist and that, above some frequency threshold, they are associated with radially 
extended power at the same frequency as the clump's orbital frequency. This intraorbital spectral 
power is significantly enhanced in the tilted datasets compared to that observed in the untilted 
simulation. Beyond identifying these variability features, however, there is little more to glean 
from this shell-averaged analysis. The next section, therefore, studies the full three-dimensional 
behavior of these structures at specific frequencies and aims to understand the variability in the 
context of the background structure discussed in Section~\ref{sec:bg}. 

\section{Structural Analysis}\label{sec:structure}

We now study the variability discussed in the previous section by analyzing its spatial structure 
and temporal behavior in more detail. In Section~\ref{subsec:mode3d}, we visualize the orbiting 
clumps and their associated intraorbital variability within the context of the flow's 
nonaxisymmetric background and analyze the temporal phase relationships across the spatial extent of 
the high-power variability. In Section~\ref{subsec:spirals}, we then put forward a unified physical 
model that explains those spectral features in each dataset.

\subsection{Spatial structure of variability}\label{subsec:mode3d}

\begin{figure*}
\centering
\def\figwd{.45\textwidth}
\begin{tabular*}{\textwidth}{@{\extracolsep{\fill}}*{2}{@{}b{\figwd}}@{}}
\setlength{\unitlength}{1in}
\figimg[\figwd]{a}{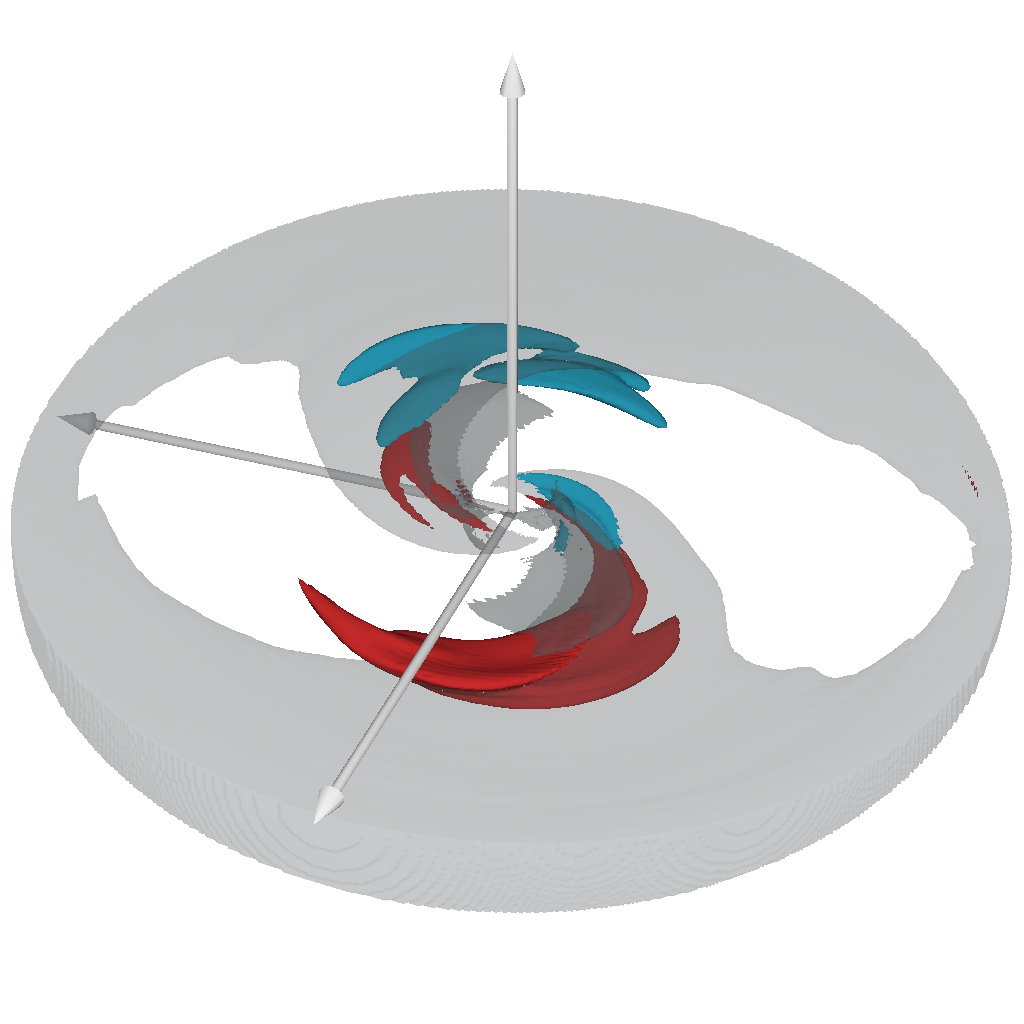} &
\figimg[\figwd]{b}{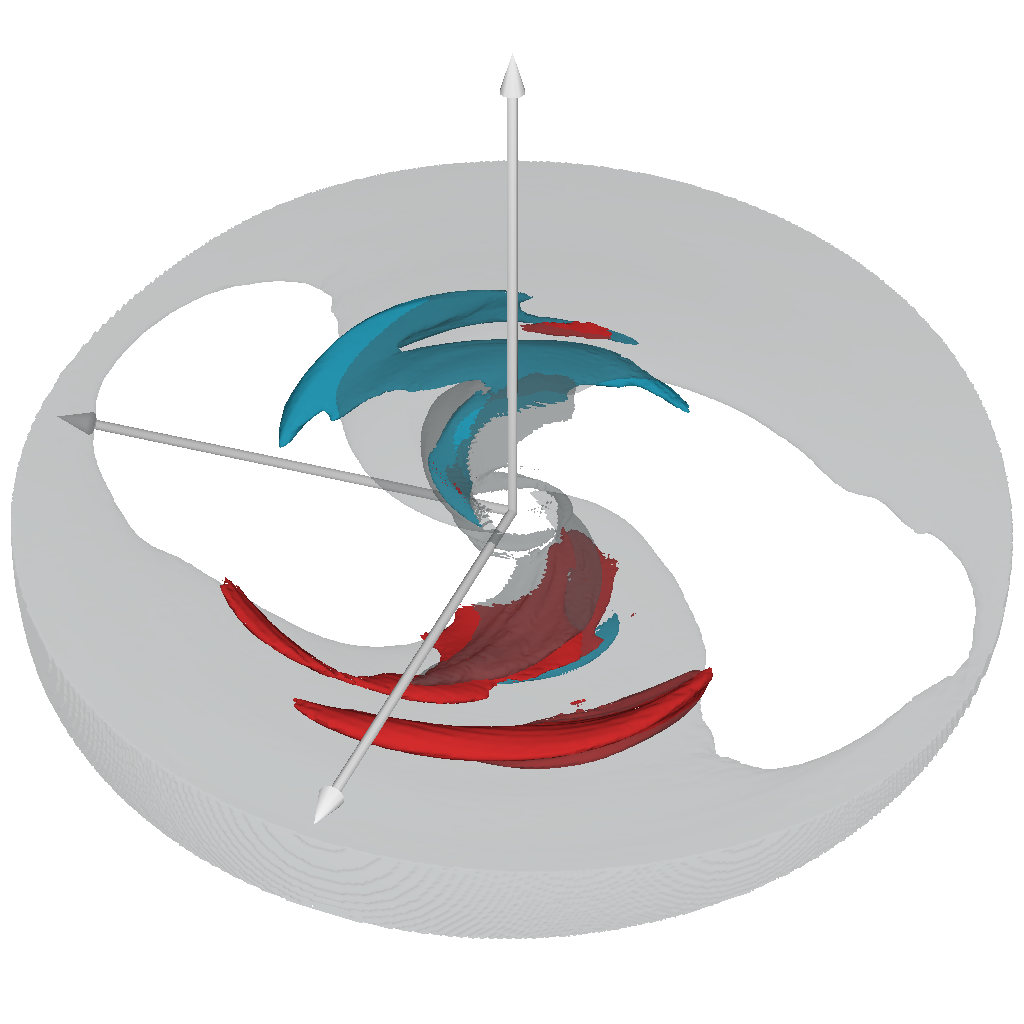} \\
\figimg[\figwd]{c}{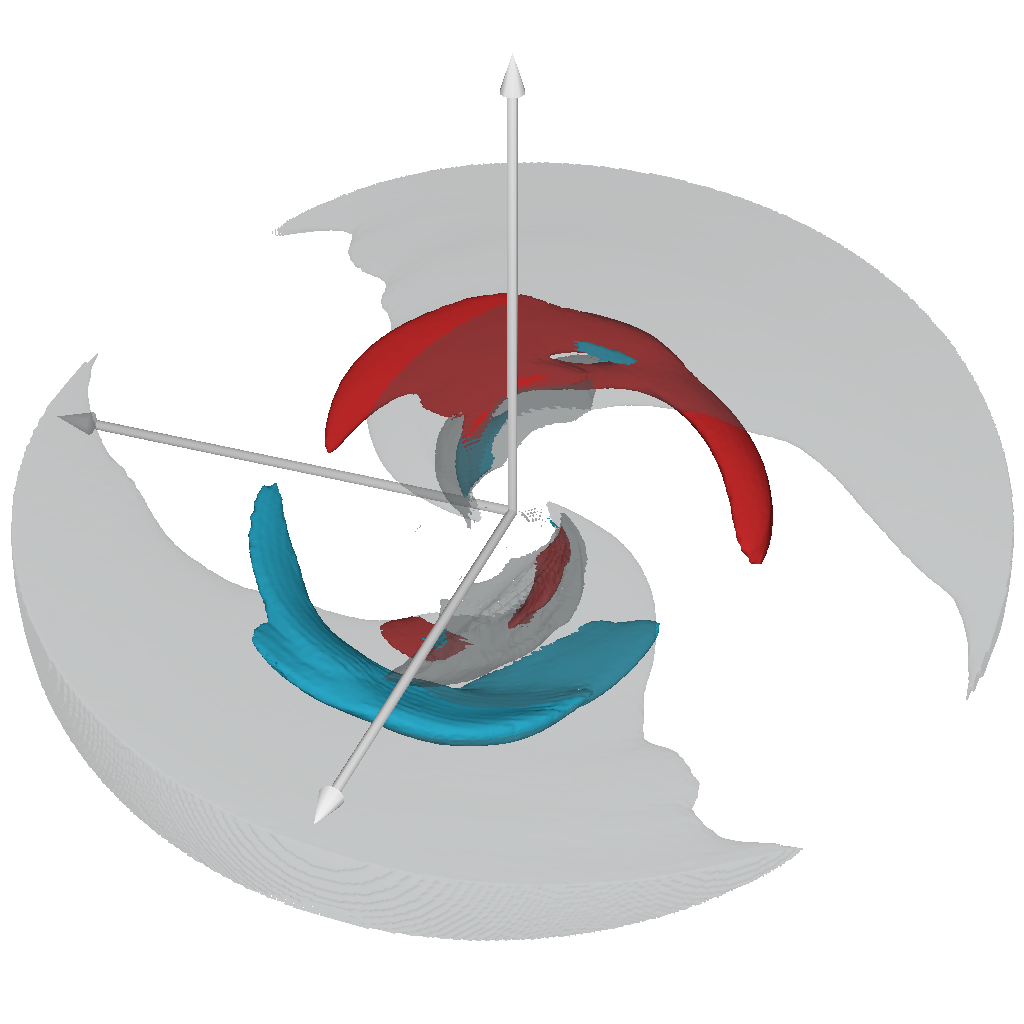} &
\figimg[\figwd]{d}{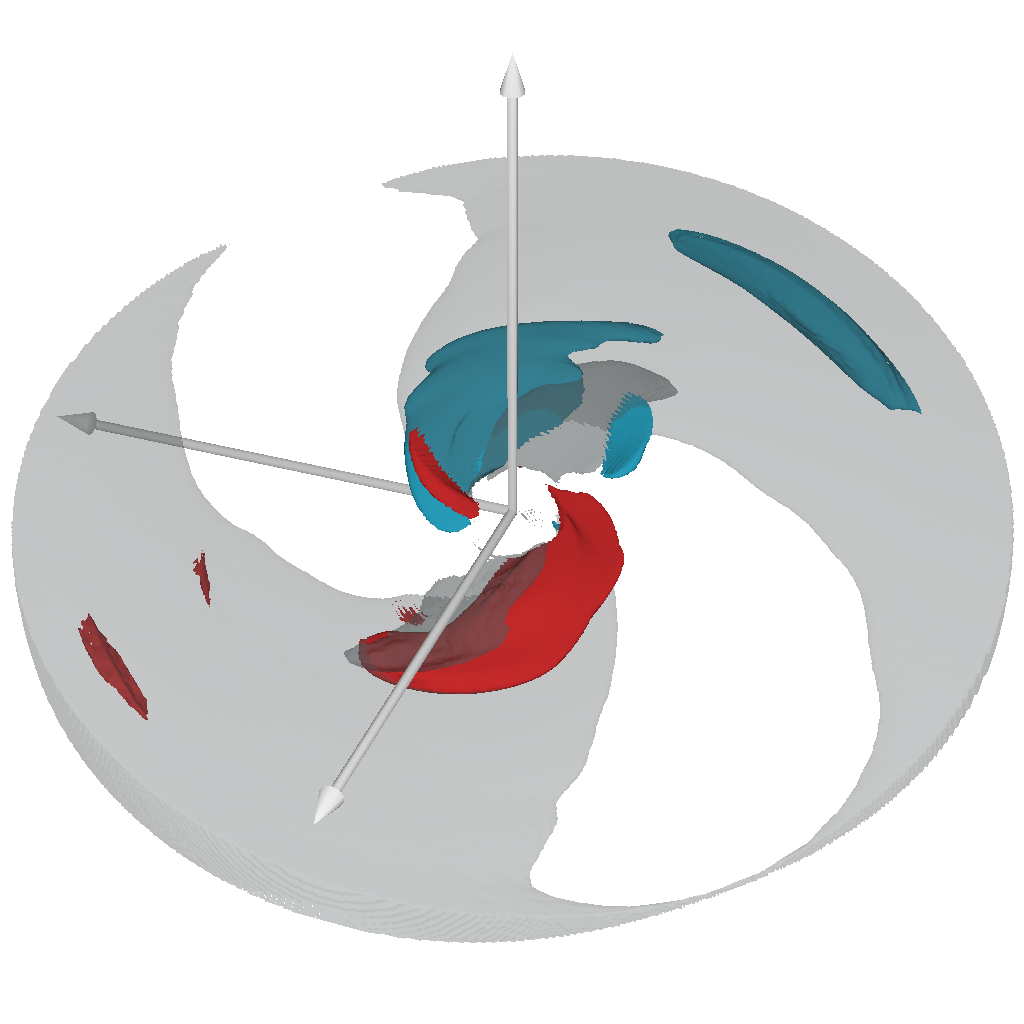}
\end{tabular*}
\caption[Three-dimensional visualizations of variability exemplifying orbiting clumps and shock
  amplified spirals.]
 {Three-dimensional visualizations of the spatial structure of the variability in density for
  oscillations at $150\Hz$ in 910h
  (a), at $118\Hz$ in 915h (b), at $140\Hz$ in 915h-64a (c), and at $60\Hz$ in 915h-64b (d).
  Specifically, we depict contours of the frequency-filtered density (see
  Equation~\ref{eq:filtered_ts}) normalized in a manner consistent with the spectra in
  Figure~\ref{fig:fp_rho}, that is,
  $\rho_\omega\sqrt{\omega/2\pi\langle\rho^2\rangle_{t,\theta,\phi}}$.  These surfaces lie at
  positive (red) and negative (blue) $0.125$ in 910h, 915h, and 915h-64a and $0.175$ in 915h-64b.
  For reference, semitransparent surfaces indicate density (light gray) and specific entropy
  generation rate (dark gray) contours identical to those in Figure~\ref{fig:shocks}. Axes extend to
  $15\RG$.
  \label{fig:fourier3d_all}}
\end{figure*}

The coherent structures identified in Section~\ref{sec:fp} represent density perturbations 
throughout the disk varying at specific frequencies. We can therefore visualize these structures by 
filtering away any fluctuations that occur on greater or smaller time scales. Mathematically, we 
construct a frequency-filtered density, 
\begin{align} 
\rho_\omega(t,\vec{r})=R_\omega\cos{\omega t}+I_\omega\sin{\omega t},
\label{eq:filtered_ts}
\end{align}
where 
$R_\omega\equiv 2{\rm Re}[\tilde\rho(\omega,\vec{r})]$ 
and 
$I_\omega\equiv 2{\rm Im}[\tilde\rho(\omega,\vec{r})]$
are twice the real and imaginary parts of the Fourier transform of the density time-series. The
factor of $2$ accounts for the fact that Fourier transforms of real-valued functions like our
density are even in frequency space, splitting total power evenly between positive and negative
frequencies. With the above definitions then, we can restrict our study to positive frequencies. This
density represents only those perturbations of the flow that vary with angular frequency $\omega$,
and therefore, although the time $t$ directly relates to the simulation coordinate time, we only
need to study the function's evolution through a single period $0\le t<2\pi/\omega$. 
Figure~\ref{fig:fourier3d_all} shows snapshots in time of contours of frequency-filtered density 
normalized in a manner consistent with the spectra in Figure~\ref{fig:fp_rho}, that is, 
$\rho_\omega\sqrt{\omega/2\pi\langle\rho^2\rangle_{t,\theta,\phi}}$. For each dataset, we have
evaluated the filtered density at frequencies that exhibit significant intraorbital power in
Figure~\ref{fig:fp_rho}. We have selected moments within the oscillation period that most clearly
illustrate that this intraorbital power is spatially concentrated near the background's over-dense
arms. The contours lie at $0.125$ for datasets  910h, 915h, and 915h-64a and $0.175$ for dataset
915h-64b.

Variability at the given frequencies in each tilted dataset spatially manifests itself in two forms: 
a clump orbiting near the corotation radius and peaking in amplitude after passing the shock fronts, 
and a pulsation within each over-dense arm just past the shocks and inside the corotation radius. 
The latter feature is responsible for the previously identified intraorbital power. 

Four explanations for this variability structure seem plausible. First, after interacting with the 
shock, the orbiting clump may, by some mechanism, liberate some small amount of material which would 
fall along the arms toward the hole. Second, after interacting with the shock, the orbiting clump 
may send a pressure wave down the over-dense arm. Third, standing inertial waves may be trapped 
within the over-dense arms and be excited and maintained by the clump-shock interaction. Fourth, a 
coherent acoustic spiral pattern produced and maintained by an orbiting clump may interact with the radially 
extended shock, increasing the amplitude of both the clump and spiral independently. 

\begin{figure*}
\centering
\setlength{\unitlength}{.36\textwidth}
\begin{tabular*}{.8\textwidth}{@{\extracolsep{\fill}}cc}
\setlength{\extrarowheight}{-128pt}
\begin{picture}(1,1)
  \put(0,0){\includegraphics[width=\unitlength]{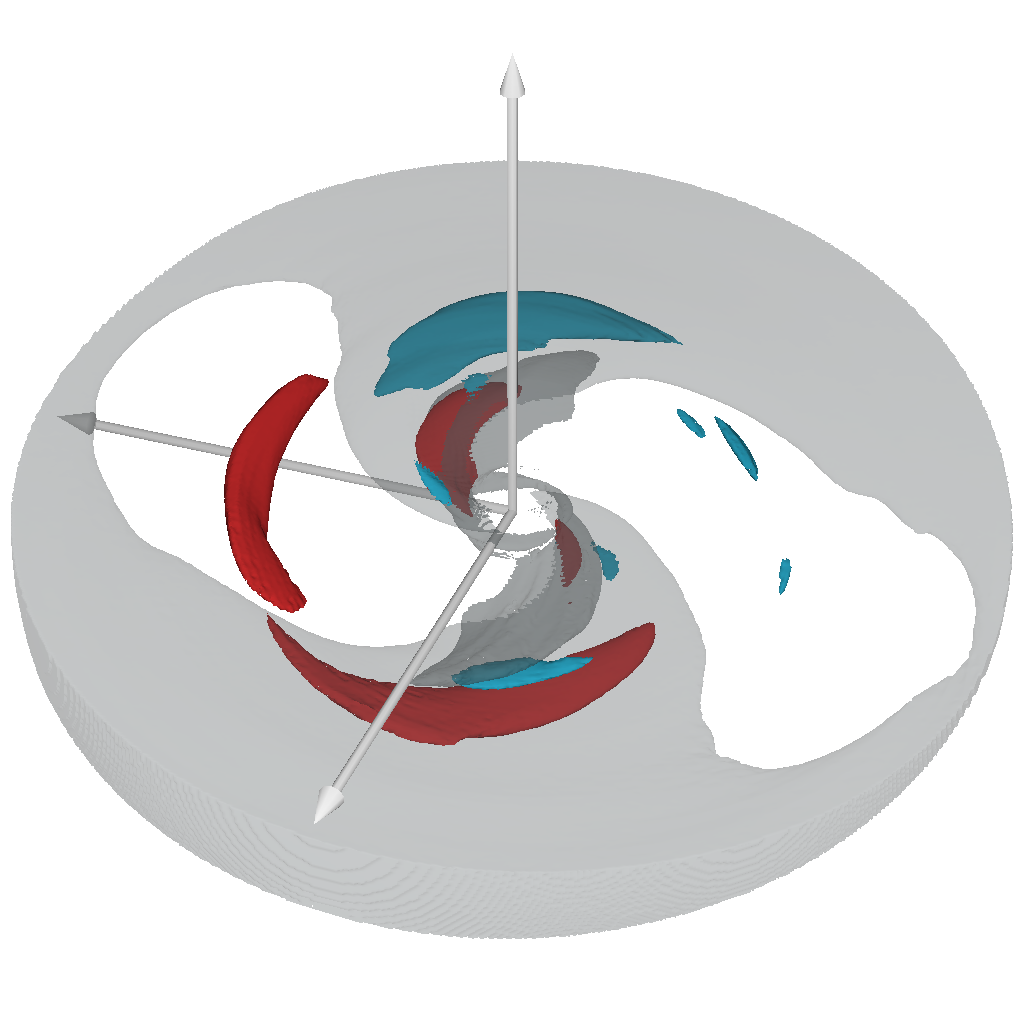}}
  \put(.5,1){\makebox(0,0)[t]{a.}}
  \put(.190,.630){\makebox(0,0)[r]{1~}\vector( 1,0){.075}}
  \put(.370,.500){\makebox(0,0)[r]{2~}\vector( 1,0){.075}}
  \put(.470,.430){\makebox(0,0)[r]{3~}\vector( 1,0){.075}}
  \put(.720,.390){\makebox(0,0)[l]{~4}\vector(-1,0){.075}}
  \put(.525,.270){\makebox(0,0)[l]{~5}\vector(-1,0){.075}}
\end{picture}
&
\begin{picture}(1,1)
  \put(0,0){\includegraphics[width=\unitlength]{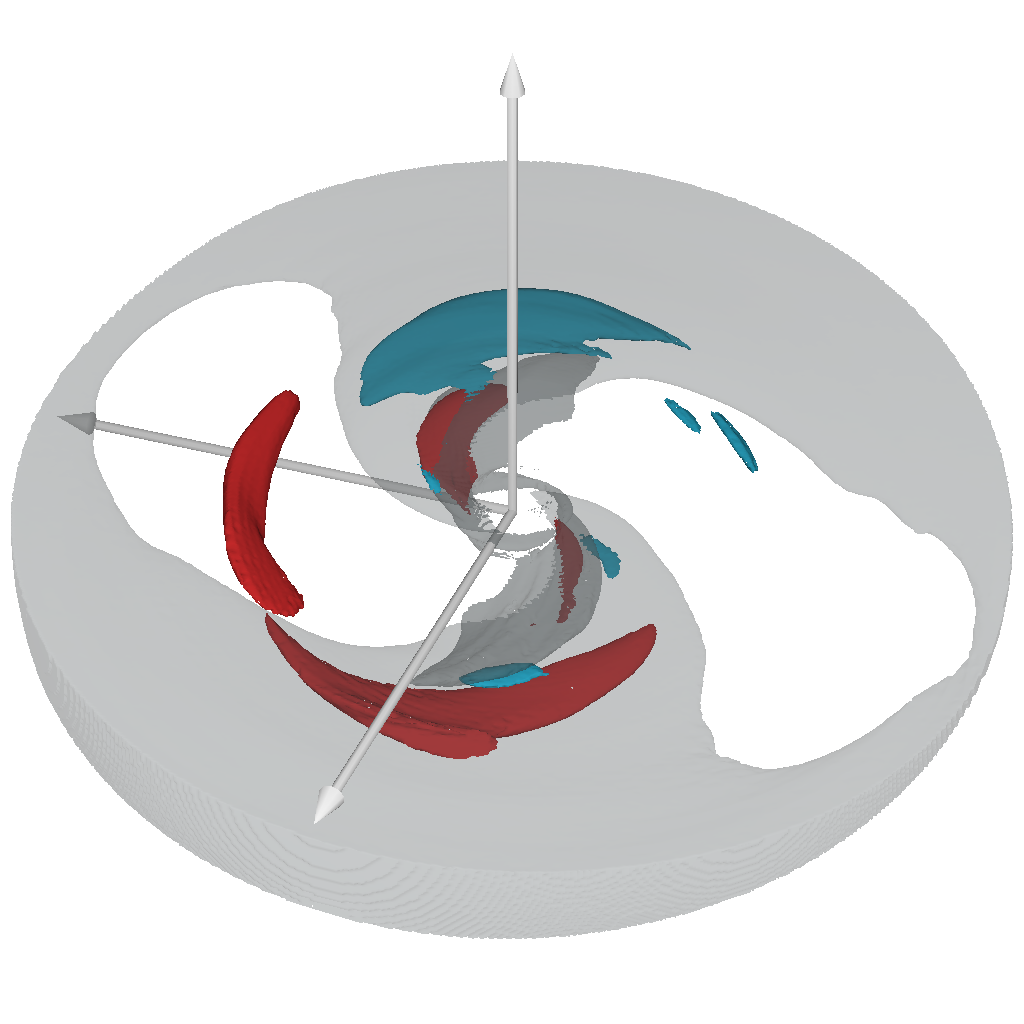}}
  \put(.5,1){\makebox(0,0)[t]{b.}}
  \put(.180,.620){\makebox(0,0)[r]{1~}\vector(1,0){.075}}
  \put(.360,.505){\makebox(0,0)[r]{2~}\vector(1,0){.075}}
  \put(.455,.390){\makebox(0,0)[r]{3~}\vector( 1,0){.075}}
  \put(.720,.390){\makebox(0,0)[l]{~4}\vector(-1,0){.075}}
  \put(.565,.270){\makebox(0,0)[l]{~5}\vector(-1,0){.075}}
\end{picture}
\\
\begin{picture}(1,1)
  \put(0,0){\includegraphics[width=\unitlength]{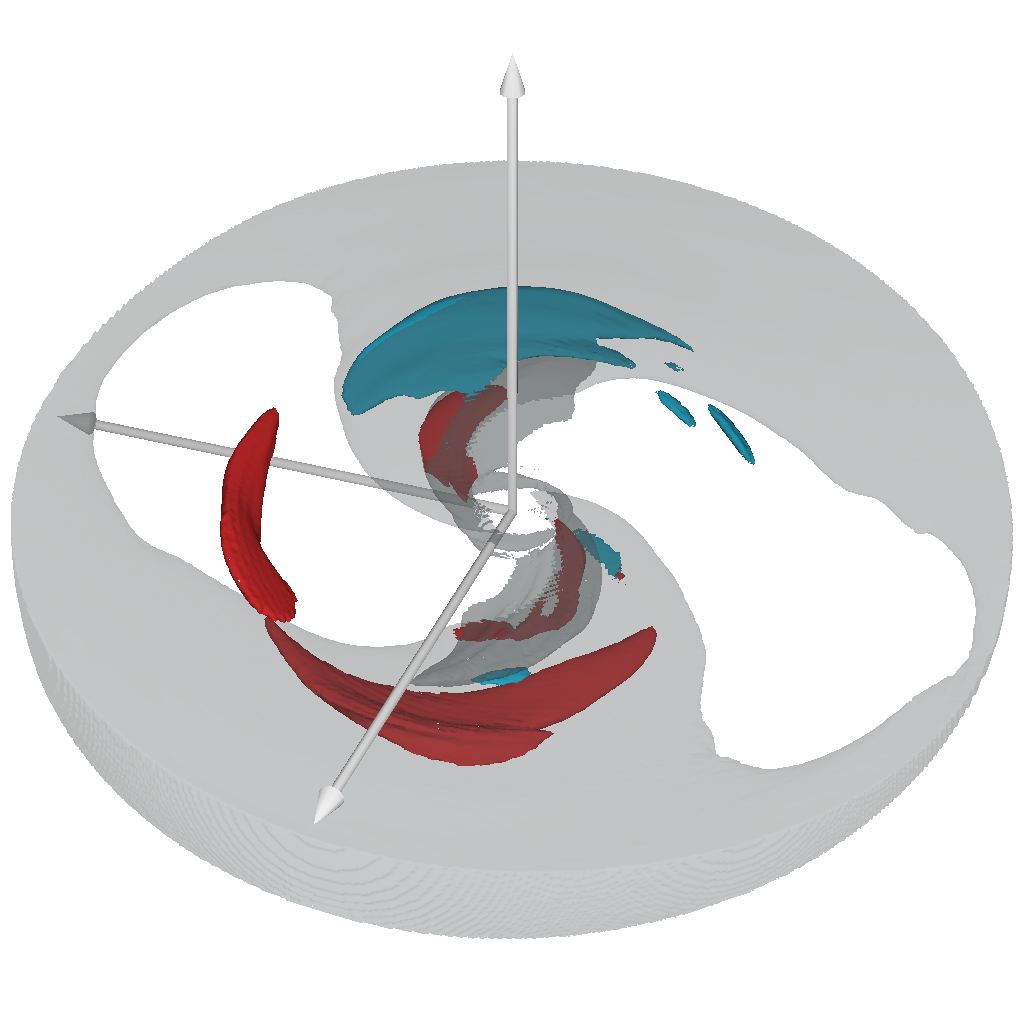}}
  \put(.5,1){\makebox(0,0)[t]{c.}}
  \put(.170,.600){\makebox(0,0)[r]{1~}\vector(1,0){.075}}
  \put(.345,.525){\makebox(0,0)[r]{2~}\vector(1,0){.075}}
  \put(.360,.380){\makebox(0,0)[r]{3~}\vector( 1,0){.075}}
  \put(.725,.385){\makebox(0,0)[l]{~4}\vector(-1,0){.075}}
  \put(.620,.280){\makebox(0,0)[l]{~5}\vector(-1,0){.075}}
\end{picture}
&
\begin{picture}(1,1)
  \put(0,0){\includegraphics[width=\unitlength]{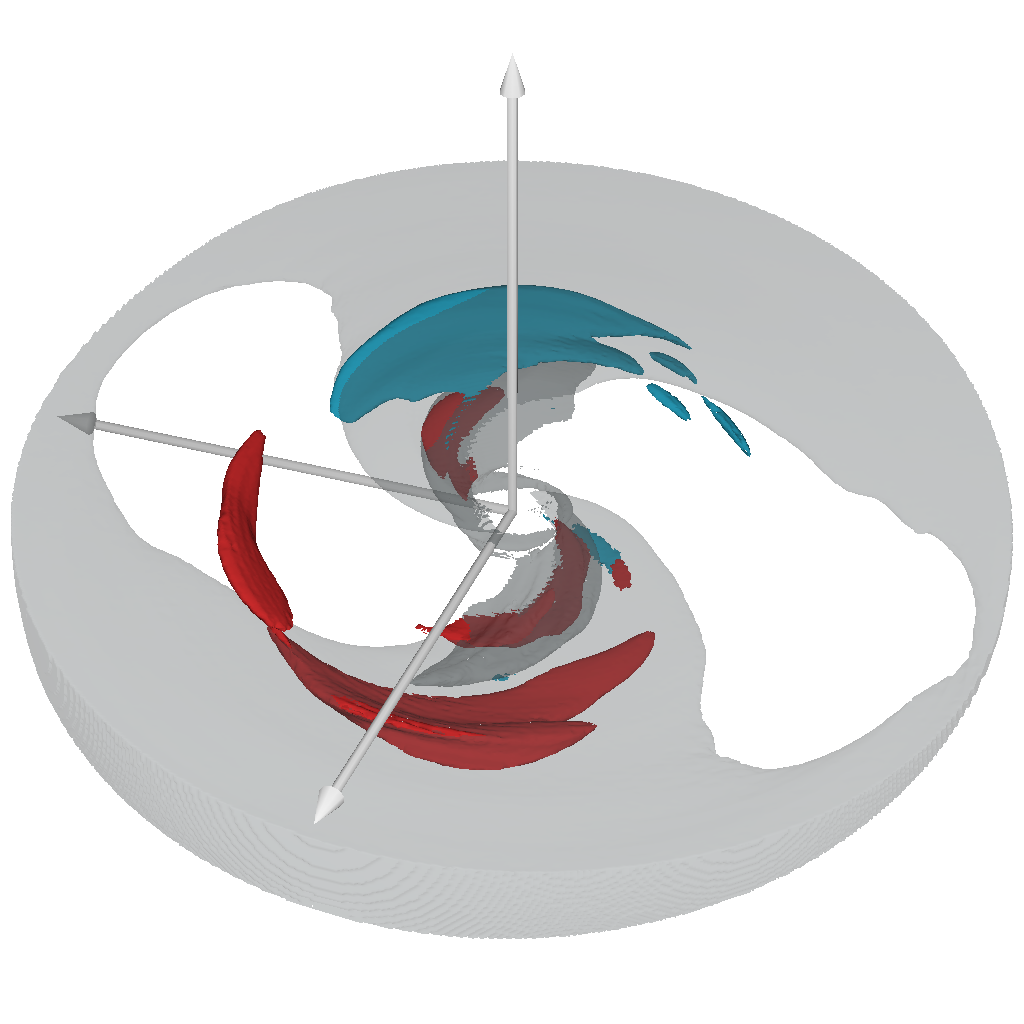}}
  \put(.5,1){\makebox(0,0)[t]{d.}}
  \put(.155,.580){\makebox(0,0)[r]{1~}\vector(1,0){.075}}
  \put(.345,.530){\makebox(0,0)[r]{2~}\vector(1,0){.075}}
  \put(.350,.385){\makebox(0,0)[r]{3~}\vector( 1,0){.050}}
  \put(.720,.380){\makebox(0,0)[l]{~4}\vector(-1,0){.075}}
  \put(.660,.290){\makebox(0,0)[l]{~5}\vector(-1,0){.075}}
\end{picture}
\\
\begin{picture}(1,1)
  \put(0,0){\includegraphics[width=\unitlength]{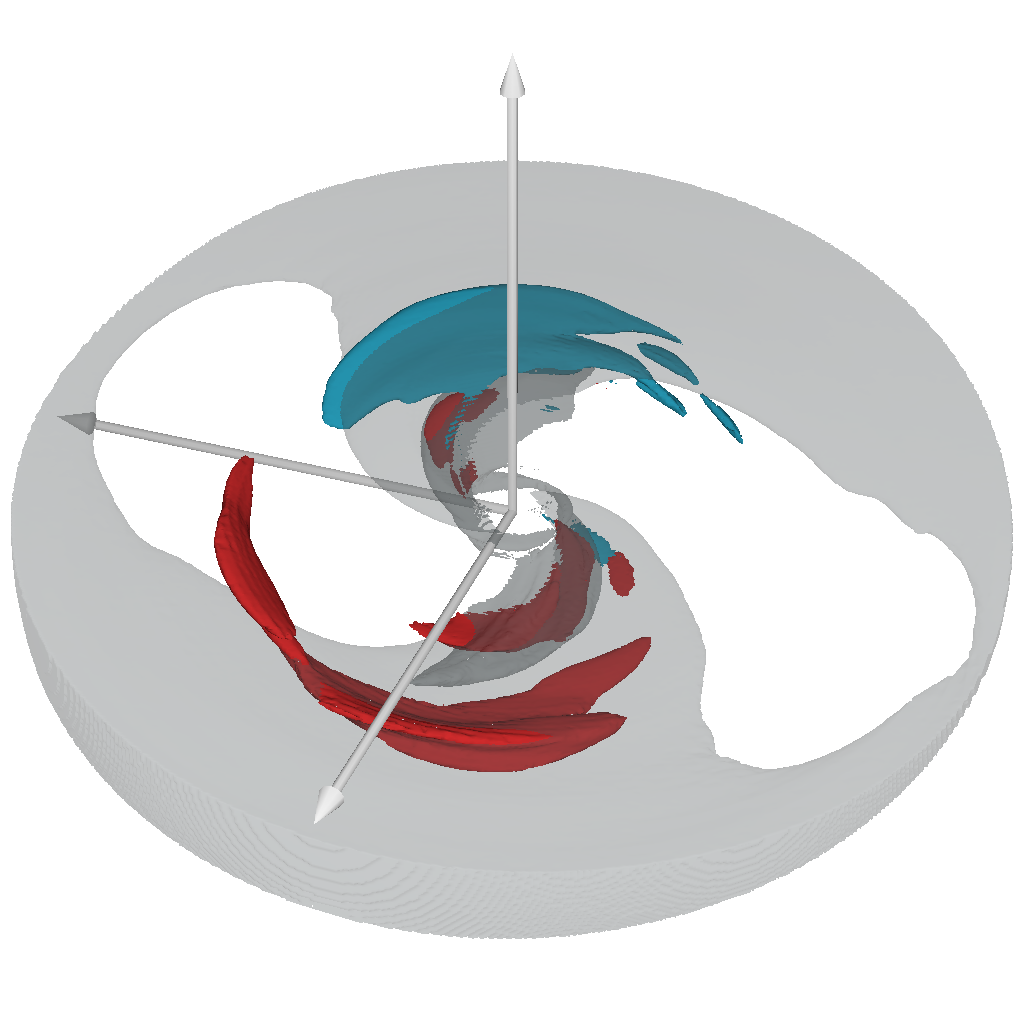}}
  \put(.5,1){\makebox(0,0)[t]{e.}}
  \put(.150,.555){\makebox(0,0)[r]{1~}\vector(1,0){.075}}
  \put(.340,.540){\makebox(0,0)[r]{2~}\vector(1,0){.075}}
  \put(.345,.390){\makebox(0,0)[r]{3~}\vector( 1,0){.050}}
  \put(.720,.375){\makebox(0,0)[l]{~4}\vector(-1,0){.075}}
  \put(.690,.295){\makebox(0,0)[l]{~5}\vector(-1,0){.075}}
\end{picture}
&
\begin{picture}(1,1)
  \put(0,0){\includegraphics[width=\unitlength]{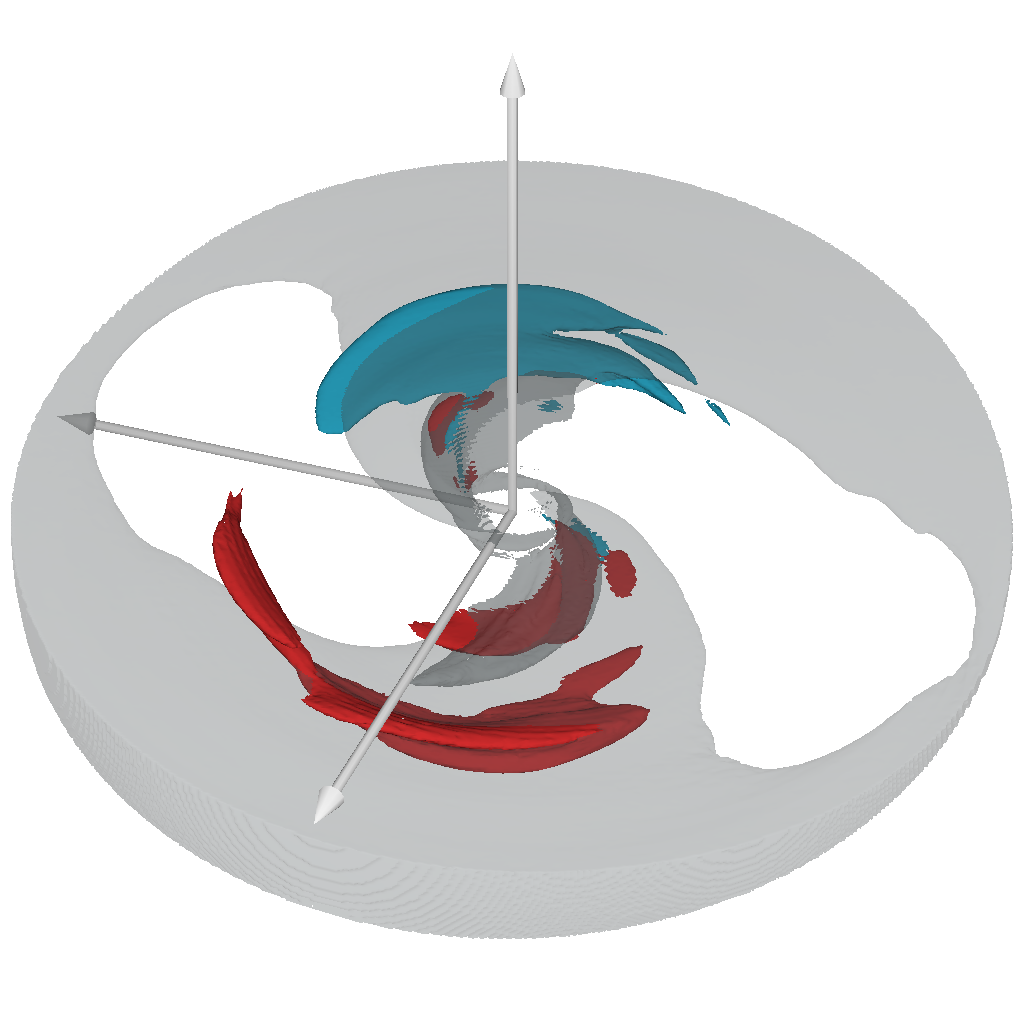}}
  \put(.5,1){\makebox(0,0)[t]{f.}}
  \put(.135,.520){\makebox(0,0)[r]{1~}\vector(1,0){.075}}
  \put(.340,.555){\makebox(0,0)[r]{2~}\vector(1,0){.075}}
  \put(.345,.390){\makebox(0,0)[r]{3~}\vector( 1,0){.045}}
  \put(.705,.365){\makebox(0,0)[l]{~4}\vector(-1,0){.075}}
  \put(.715,.305){\makebox(0,0)[l]{~5}\vector(-1,0){.075}}
\end{picture}
\end{tabular*}
\caption[Successive three-dimensional snapshots of variability in 915h at $118\Hz$
  highlighting apparently outward moving flashes.]
 {Successive three-dimensional snapshots in time of the $118\Hz$ pattern in the 915h simulation
  given by normalized frequency-filtered density (see Figure~\ref{fig:fourier3d_all}). By
  construction, this variability is independent of any local characteristic frequencies like the
  orbital frequency.
  Each frame is separated by $1/40$th of a full oscillation period. Semi-transparent
  surfaces (light and dark gray) and positive (red) and negative (blue) contours lie at the same
  levels as in Figure~\ref{fig:fourier3d_all}. Numbered arrows track the location of the trailing
  ($1$) and leading ($5$) edges of the orbiting clump as well as three edges of the flashes in the
  over-dense arms: one which advances outwardly from the hole ($3$) and two that recede away from
  it ($2$ and $4$). Axes extend to $15\RG$.
  \label{fig:fourier3d_915h}}
\end{figure*}
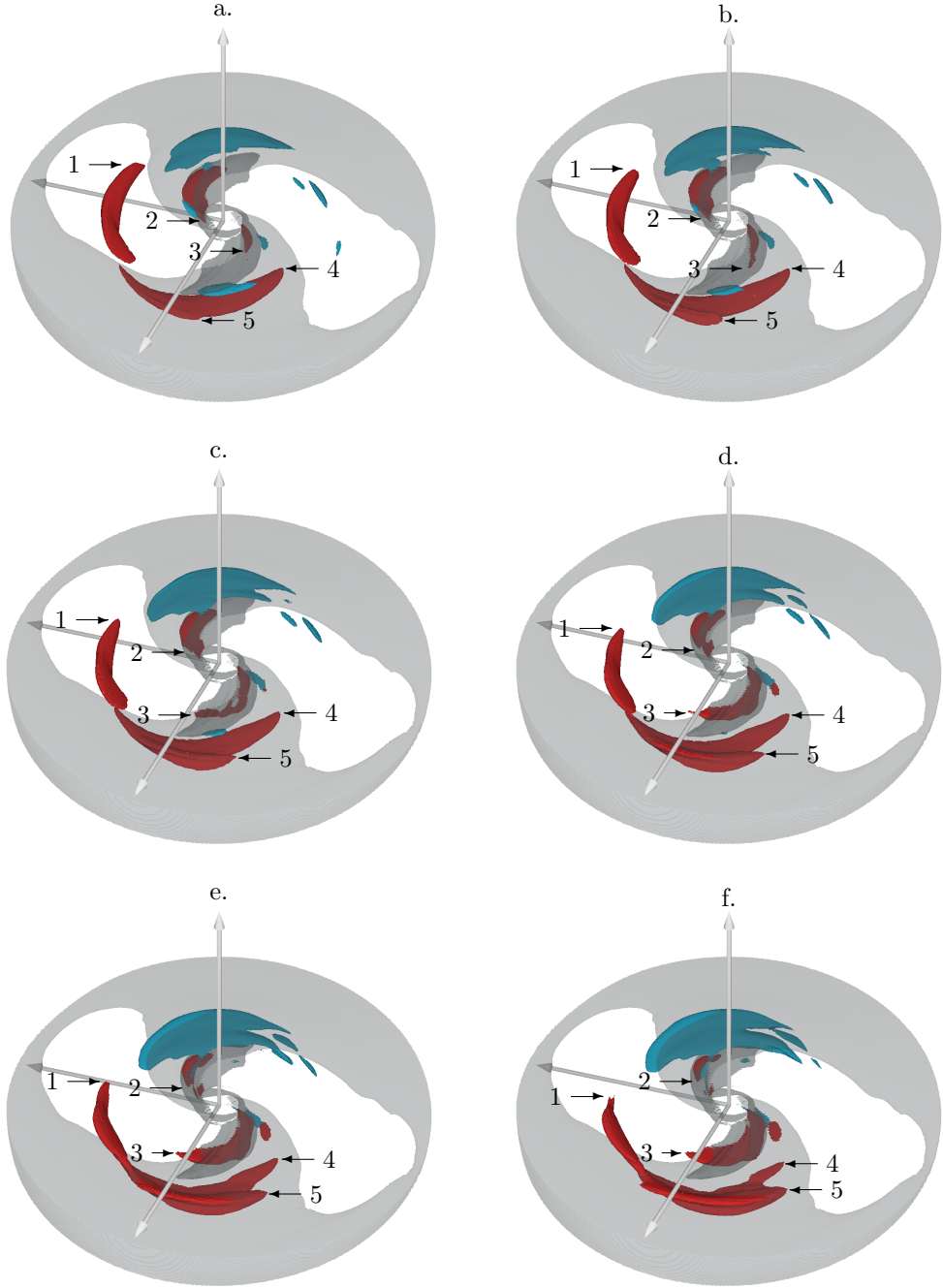

In light of our conclusions in Section~\ref{sec:bg}, we rule out the first explanation since 
sloughed off material cannot simply fall past the otherwise roughly Keplerian flow circulating 
around the hole. Similarly, a close examination of the temporal phase relationships within the 
pulses forces us to discard the second and third explanations. Specifically, 
Figure~\ref{fig:fourier3d_915h} shows six successive snapshots in time of the $118\Hz$ variability, 
that is, of the frequency-filtered density given by Equation~\ref{eq:filtered_ts}. The time interval 
between each frame is $1/40$th of the oscillation's period. By construction, these structures vary 
at the filter frequency, in this case, $118\Hz$, regardless of any local characteristic frequencies 
like the orbital frequency. Notice that the red variability structure inside the over-dense arm 
closest to the viewer (arrow $3$) appears to start at small radius and moves outward toward larger 
radii. Likewise, two additional structures (labeled $2$ and $4$) appear to recede away from the 
center. Since the second and third hypotheses predict an inwardly propagating wave and a standing 
wave within the arms, respectively, this behavior rules out both models. 

In the fourth hypothesis however, a radially extended spiral pattern rotates bodily about the
hole at the same frequency (here $118\Hz$) as its associated corotating clump. Depending on the
shape of the spiral relative to a shock surface, the point of interaction between the two
may appear to move inward or outward in time along the shock's face. For example, if some segment of
the spiral is trailing, the interaction point between it and the stationary shock surface will
appear to move outward in time, akin to structure~$3$ in Figure~\ref{fig:fourier3d_915h}. In
this way, a spiral-shock interaction model permits all the complex temporal relationships depicted
in Figure~\ref{fig:fourier3d_915h}. Of our four hypotheses, this remains the only plausible
explanation.

At the spiral-shock interaction point, the acoustic wave passing through the shock will
experience an enhancement in its density 
amplitude (as defined by half the difference between the maximum and minimum density in the wave) 
equal to the shock compression ratio discussed in Section~\ref{sec:bg}. In the $15^\circ$ tilted 
simulations, the ratios were as high as $4.7$ while in the $10^\circ$ tilted simulation they were as 
large as $4.0$ \citep{gen12}. We therefore expect density power in the acoustic waves (which goes as the square of 
the amplitude) to be enhanced by a factor of roughly $22$ in the $15^\circ$ tilted simulation, and 
$16$ in the $10^\circ$ tilted simulations, as compared to the untilted simulation. This 
quantitatively matches the larger intraorbital power in the vertical streaks in the tilted 
simulations as compared to the untilted simulation in the density power spectra shown in 
Figure~\ref{fig:fp_rho}. This spiral-shock interaction model can explain both the apparent motion
of the variability along the background over-dense arms and the magnitude of the increased
variability at the point of interaction.

\subsection{Prograde and retrograde decomposition}\label{subsec:spirals}
Of those models described above that aim to explain the apparent movement of perturbations along the
over-dense arms, only that in which a rotating acoustic spiral accompanies the orbiting clumps 
identified in Section~\ref{sec:fp} and visualized in section \ref{subsec:mode3d} is consistent with 
the observed temporal relationships. In an axisymmetric background akin to that in our untilted 
simulation, for example, such a spiral pattern should be both stationary in a frame corotating with 
the clump and largely $m=1$ in its azimuthal symmetry. By assuming these characteristics, this 
section further decomposes the variability in our simulations to locate these spiral structures and 
better understand their physical nature.

Since the structures visualized in the previous section are largely $m=1$,
we first isolate this component of the variability from the frequency-filtered time series in
Equation~\ref{eq:filtered_ts}. In other words, we project the spatial functions $R_\omega$ and
$I_\omega$ onto the two $m=1$ spatial basis functions so that
\begin{align}
\rho_{\omega,m=1}&=(R_{\omega,\c}\cos{\phi}+R_{\omega,\s}\sin{\phi})\cos{\omega t} \nonumber\\
&\qquad+(I_{\omega,\c}\cos{\phi}+I_{\omega,\s}\sin{\phi})\sin{\omega t},
\end{align}
where
\begin{subequations}
\begin{align}
R_{\omega,\c}&=\frac{1}{\pi}\int_0^{2\pi}R_\omega\cos{\phi}\,d\phi \\
R_{\omega,\s}&=\frac{1}{\pi}\int_0^{2\pi}R_\omega\sin{\phi}\,d\phi \\
I_{\omega,\c}&=\frac{1}{\pi}\int_0^{2\pi}I_\omega\cos{\phi}\,d\phi \\
I_{\omega,\s}&=\frac{1}{\pi}\int_0^{2\pi}I_\omega\sin{\phi}\,d\phi.
\end{align}
\end{subequations}
Using basic trigonometric identities, one can combine these terms in a straight-forward manner to find
\begin{align}
\rho_{\omega,m=1}&= A_\omega\cos(\phi-\omega t-\alpha_\omega)  \nonumber\\
&\qquad+ B_\omega\cos(-\phi-\omega t-\beta_\omega),
\label{eq:pro_ret}
\end{align}
where
\begin{subequations}
\begin{align}
A_\omega^2&=\frac{1}{4}\left[\left(R_{\omega,\c}+I_{\omega,\s}\right)^2\right. \nonumber\\
               &\qquad\left.+\left(R_{\omega,\s}-I_{\omega,\c}\right)^2\right] \\
B_\omega^2&=\frac{1}{4}\left[\left(R_{\omega,\c}-I_{\omega,\s}\right)^2\right. \nonumber\\
               &\qquad\left.+\left(R_{\omega,\s}+I_{\omega,\c}\right)^2\right] \\
\tan\alpha_\omega&=\frac{ R_{\omega,\s}-I_{\omega,\c}}{R_{\omega,\c}+I_{\omega,\s}} \\
\tan\beta_\omega&=\frac{-R_{\omega,\s}-I_{\omega,\c}}{R_{\omega,\c}-I_{\omega,\s}}.
\end{align}
\label{eq:pro_ret_funcs}
\end{subequations}
Given the form of the arguments within each cosine, we can now interpret the coefficients $A_\omega$ 
and $B_\omega$ as the amplitudes of the two rotating components, a prograde pattern and a retrograde 
pattern, respectively, at each $(r, \theta)$ pair. 

Four basic motions can result from this interpretation on a given ring of constant $r$ and $\theta$. 
Clearly, if $A_\omega$ or $B_\omega$ is zero, one would observe a purely retrograde or prograde 
motion, respectively. Similarly, $A_\omega=B_\omega$ indicates a standing wave structure. Less 
intuitively, the more general case in which $0<B_\omega<A_\omega$ turns out to be a prograde pattern 
that no longer exhibits a constant amplitude or rotation speed. Like two sinusoidal waves of 
differing amplitudes crossing on a string, the two components constructively interfere at some
moments producing a large density perturbation (of magnitude $|A_\omega|+|B_\omega|$) that rotates 
in a prograde manner at an instantaneous angular speed smaller than the angular frequency of the 
oscillation. In the same way, at the moments when the two components destructively interfere, they 
produce a lower amplitude perturbation (having magnitude $|A_\omega|-|B_\omega|$) that again rotates 
in a prograde fashion but now with an instantaneous angular speed greater than the angular frequency 
of the oscillation. 

Keeping these possible motions in mind, we return to our untilted and tilted disks. In the 
axisymmetric background of the untilted simulation, the decomposition of a clump with a corotating 
spiral ought to return a purely prograde pattern. On the other hand, we suggested in 
Section~\ref{subsec:mode3d} that the standing shocks associated with the non-axisymmetric background 
amplify the density of the clump and its corotating spiral in the tilted simulations. We therefore 
expect to find a predominantly prograde spiral throughout the region interior to the clump 
accompanied by a weaker but nonzero retrograde component. This latter piece merely accounts for the 
fact that the complete perturbation has a higher magnitude and slower rotation speed just past the 
standing shocks and a lower amplitude and faster rotation speed away from the shocks. This result 
aligns well with the structure and evolution depicted in Figures~\ref{fig:fourier3d_all} and 
\ref{fig:fourier3d_915h}, respectively. 

\begin{figure*}
\centering
\begin{tabular*}{\textwidth}{@{\extracolsep{\fill}}*{2}{@{}b{\figwd}}@{}}
\figimg{a}{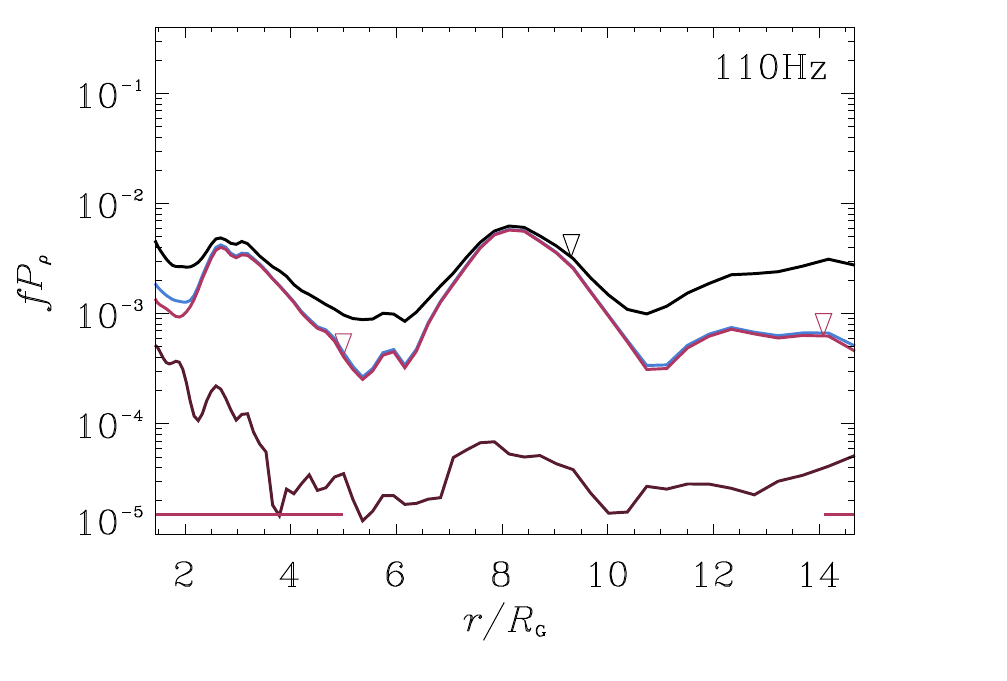} &
\figimg{b}{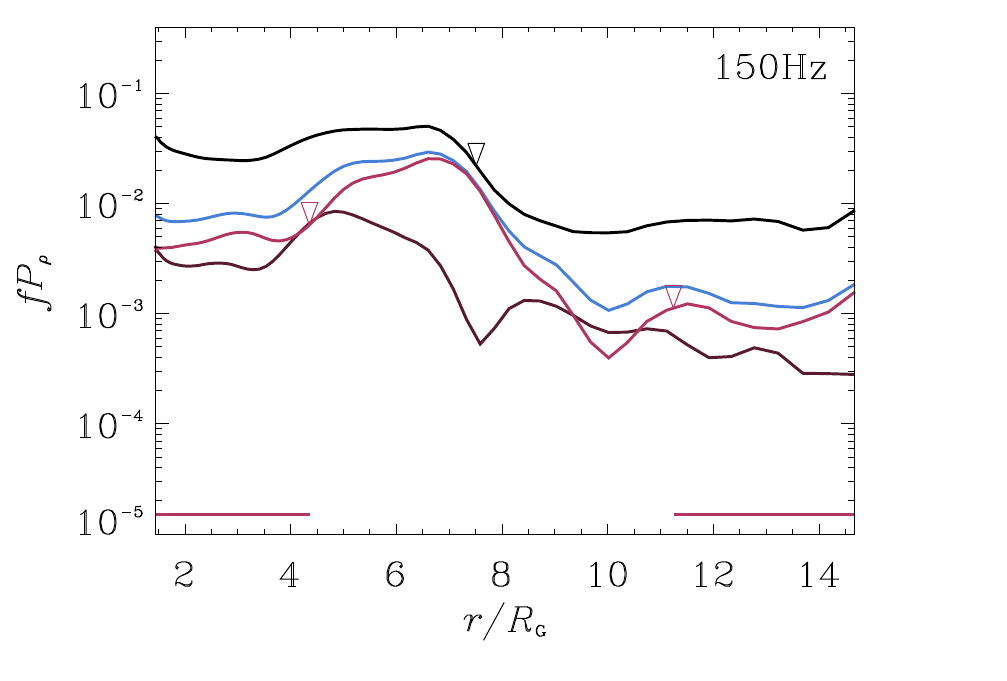} \\
\figimg{c}{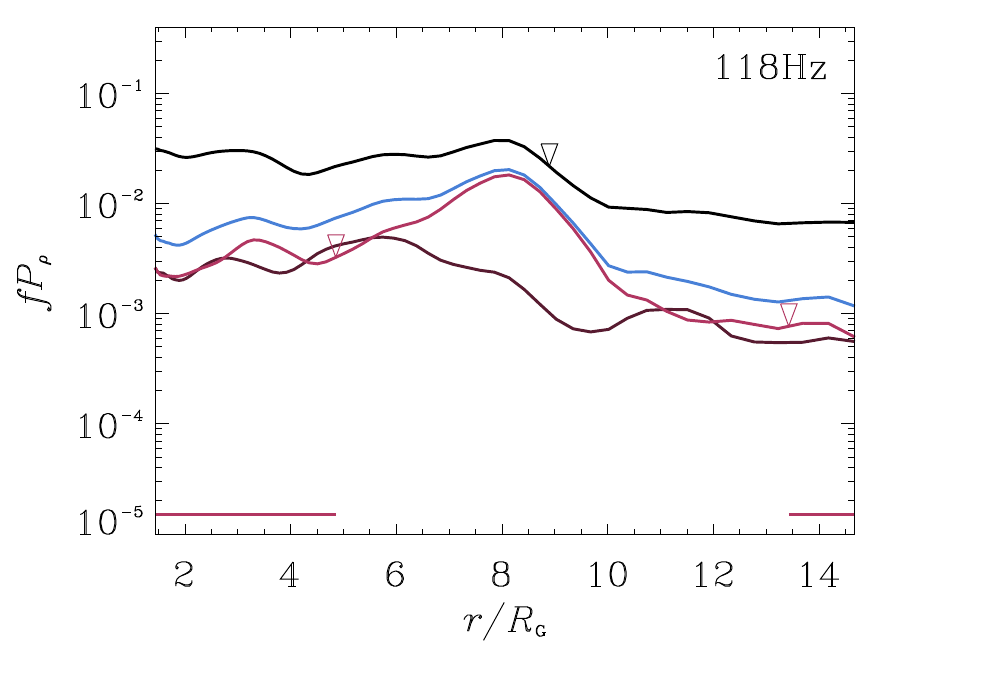} &
\figimg{d}{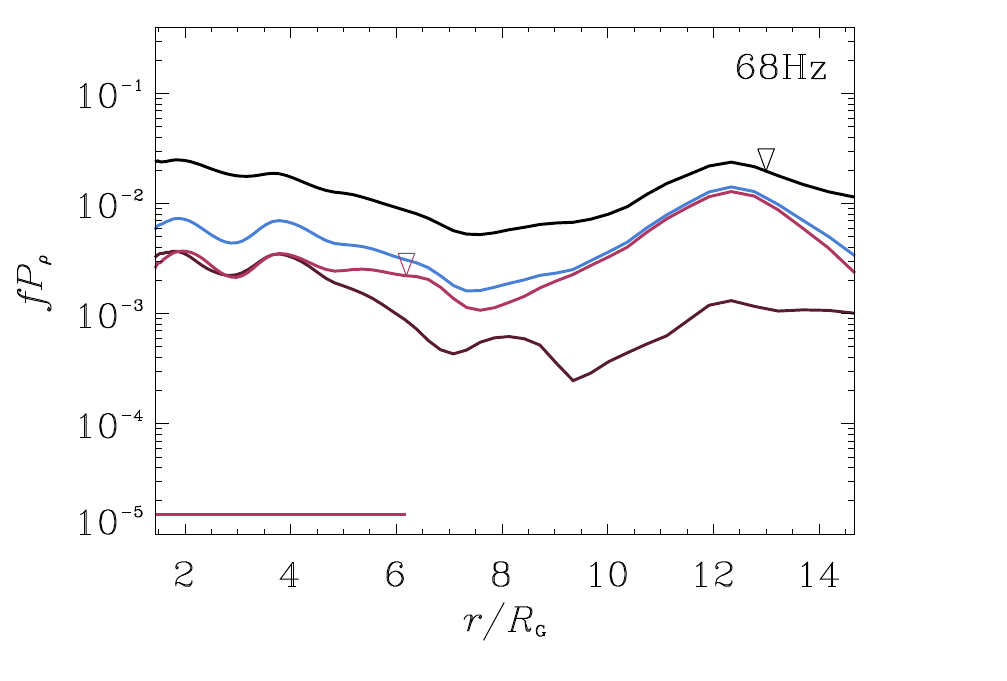} \\
\figimg{e}{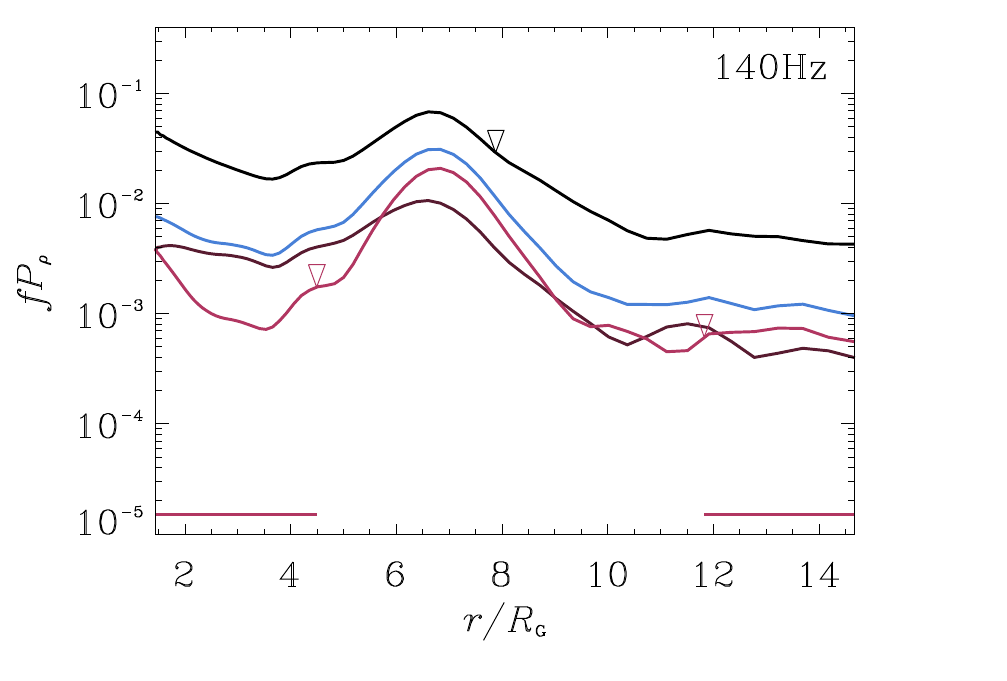} &
\figimg{f}{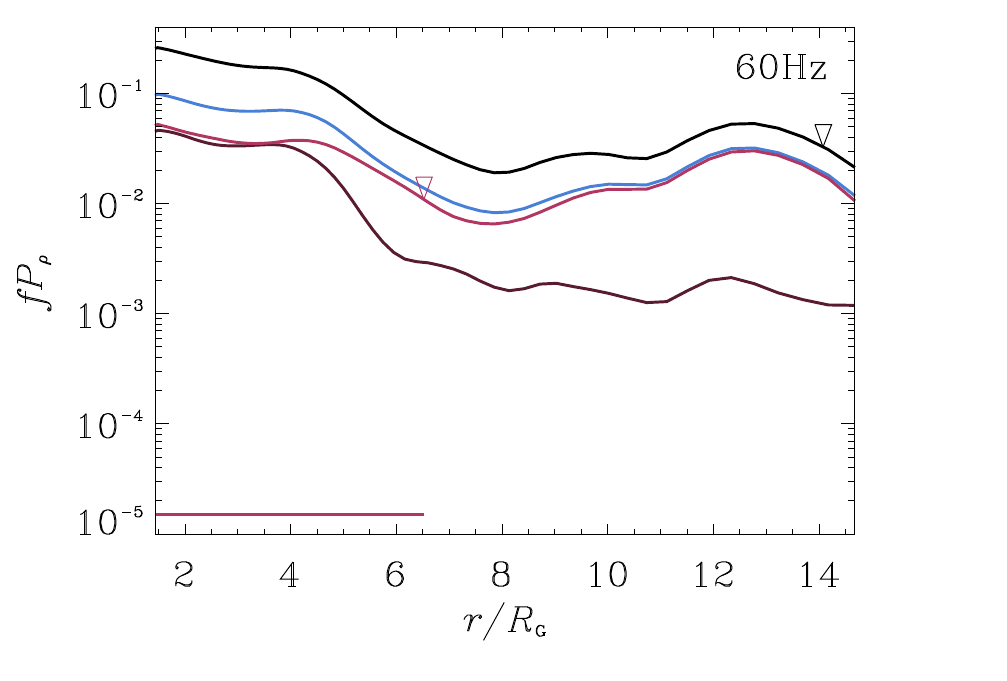}
\end{tabular*}
\caption[Decomposition of the $m=1$ variability in density into its prograde and retrograde
  contributions.]
 {Total power at specified frequencies (black; $\langle R_\omega^2+I_\omega^2\rangle_{\theta,\phi}$)
  in 90h (a), 910h (b), 915h (c and d), 915h-64a (e), and 915h-64b (f) compared with the
  power in the $m=1$ projection at the indicated frequencies (blue;
  $\langle A_\omega^2+B_\omega^2\rangle_\theta$). The projected power is further decomposed into its
  prograde (light red; $\langle A_\omega^2\rangle_\theta$) and retrograde (dark red;
  $\langle B_\omega^2\rangle_\theta$) contributions. Note that all curves have been multiplied by
  $\omega/2\pi\langle\rho^2\rangle_{t,\theta,\phi}$ to match the normalization of the spectra in
  Figure~\ref{fig:fp_rho}. Triangles mark the test particle
  corotation radius (black) and $m=1$ Lindblad resonances (light red). Light red, horizontal lines
  mark propagation regions for $n=0$, $m=1$ modes.
  \label{fig:mode_decomp}}
\end{figure*}

We plot in Figure~\ref{fig:mode_decomp} the power in each rotating component, 
$\langle|A_\omega|^2\rangle_\theta$ and $\langle|B_\omega|^2\rangle_\theta$ at the indicated 
frequencies and as a function of radius, normalizing to match Figure~\ref{fig:fp_rho}. As 
anticipated, the $m=1$ variability in the untilted simulation is dominated by a prograde pattern at 
all radii and shows very little power in the retrograde term. We take this as evidence that the weak 
intraorbital variability found in the power spectra in Figure~\ref{fig:fp_rho} for the untilted 
simulation arises from prograde spirals corotating with Keplerian clumps. 

For the tilted simulations, Figure~\ref{fig:mode_decomp} shows that the clump is likewise dominated 
by prograde motion. We interpret the prograde pattern as, in some sense, the average power profile 
of the clump's associated spiral. Additionally though, we also see an enhanced retrograde component 
and interpret the relative retrograde to prograde power as the degree to which the shock amplifying 
background deforms the spiral as it rotates around the disk. We emphasize that one must only 
interpret the retrograde power as a mathematical tool; it cannot be any kind of physical structure. 
If it were a real, propagating perturbation, it would be traveling upstream, against the background 
fluid flow at speeds which far exceed the local sound speed. Nonetheless, the largely prograde 
nature of the variability corroborates the clump-spiral hypothesis in both untilted and tilted 
configurations and further indicates that the standing shocks associated with tilted geometries 
enhance variability by effectively amplifying passing density fluctuations. 

\begin{figure*}
\centering
\begin{tabular*}{\textwidth}{@{\extracolsep{\fill}}*{2}{@{}b{\figwd}}@{}}
\figimg{a}{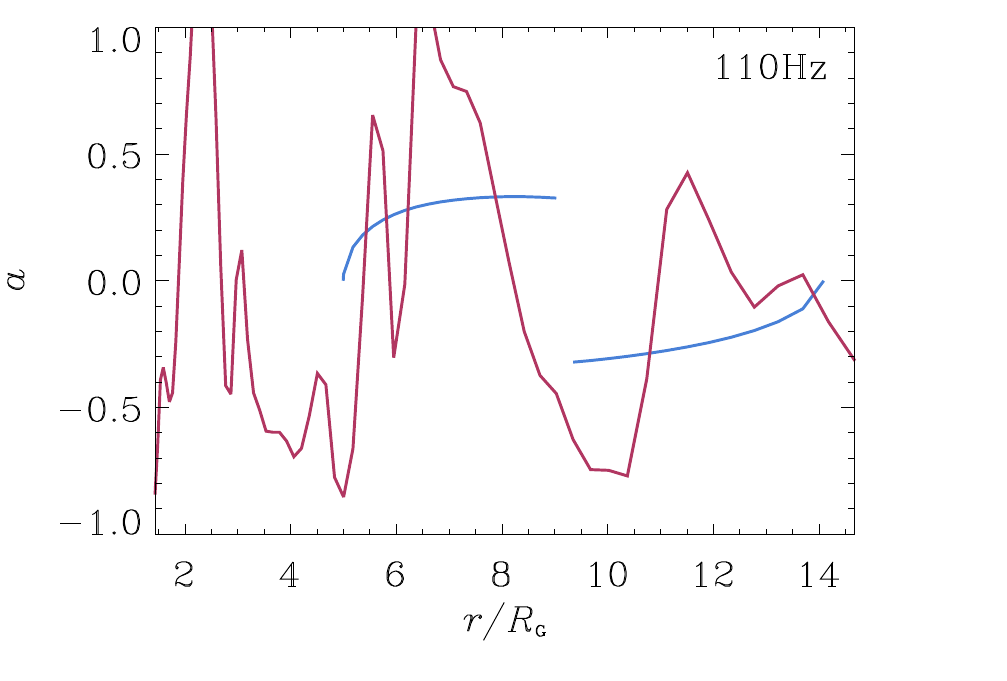} &
\figimg{b}{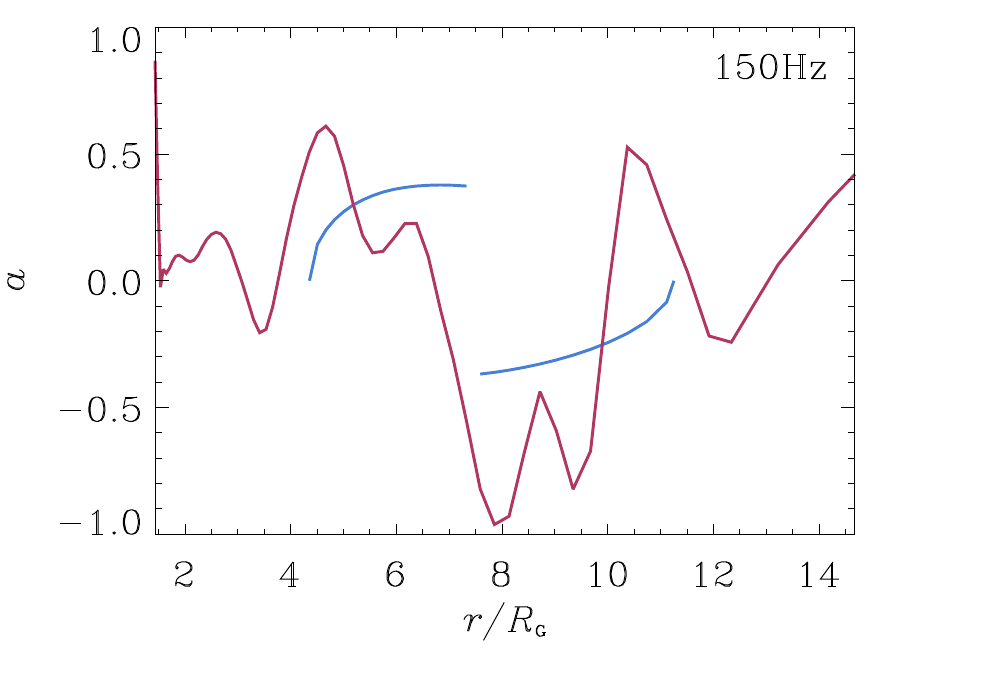} \\
\figimg{c}{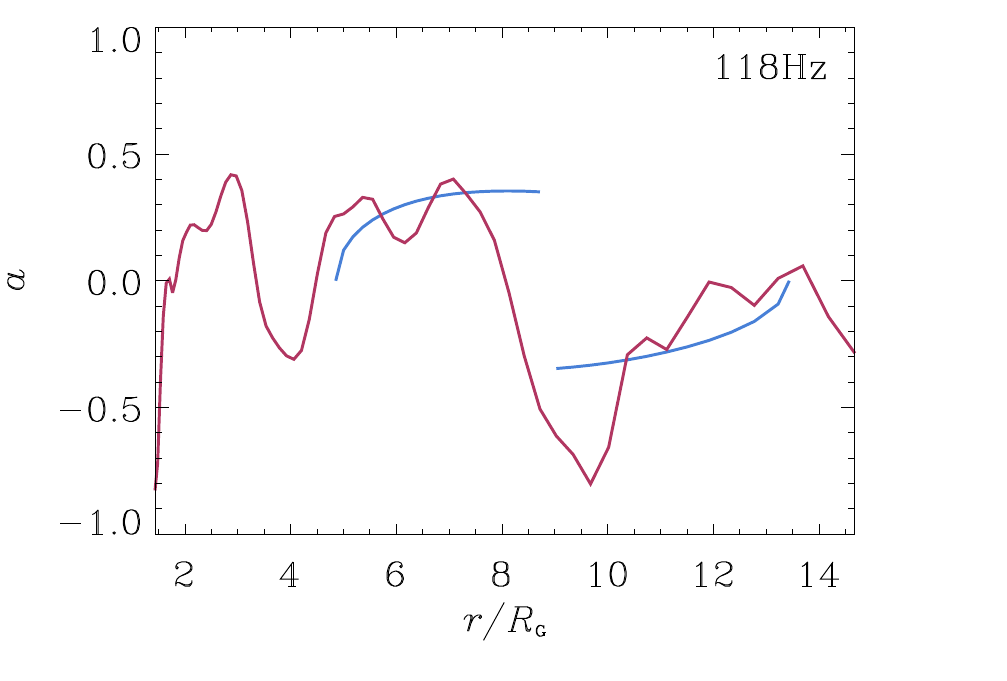} &
\figimg{d}{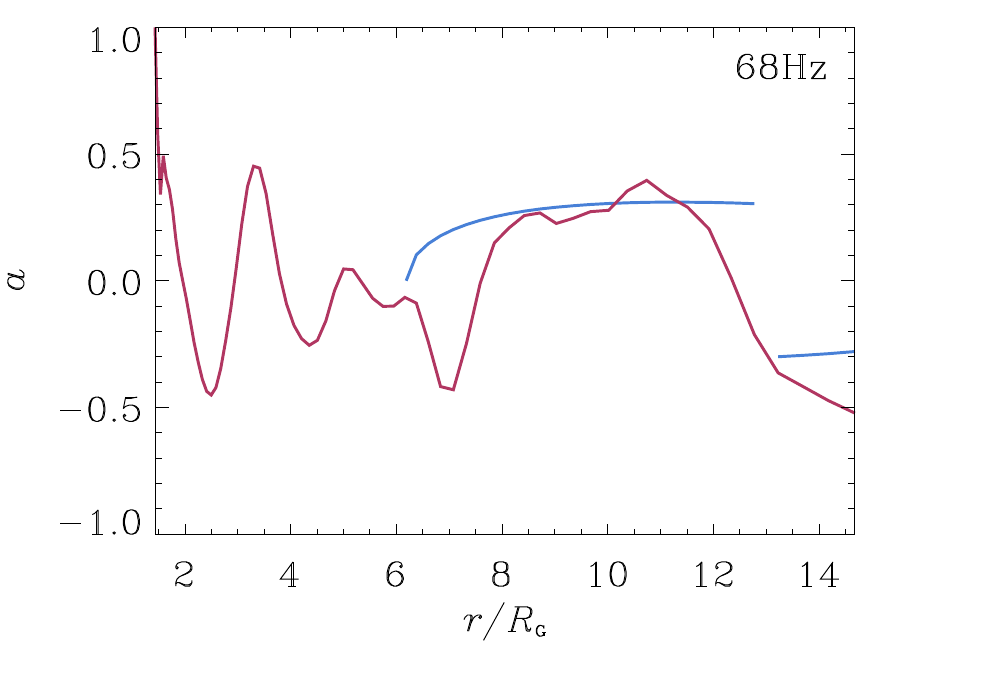} \\
\figimg{e}{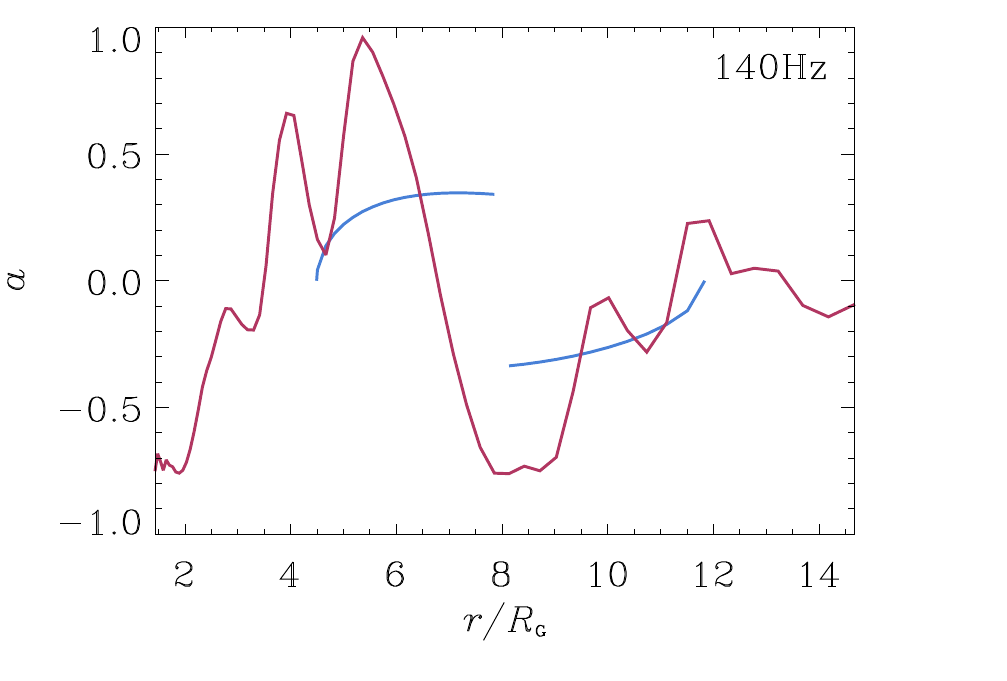} &
\figimg{f}{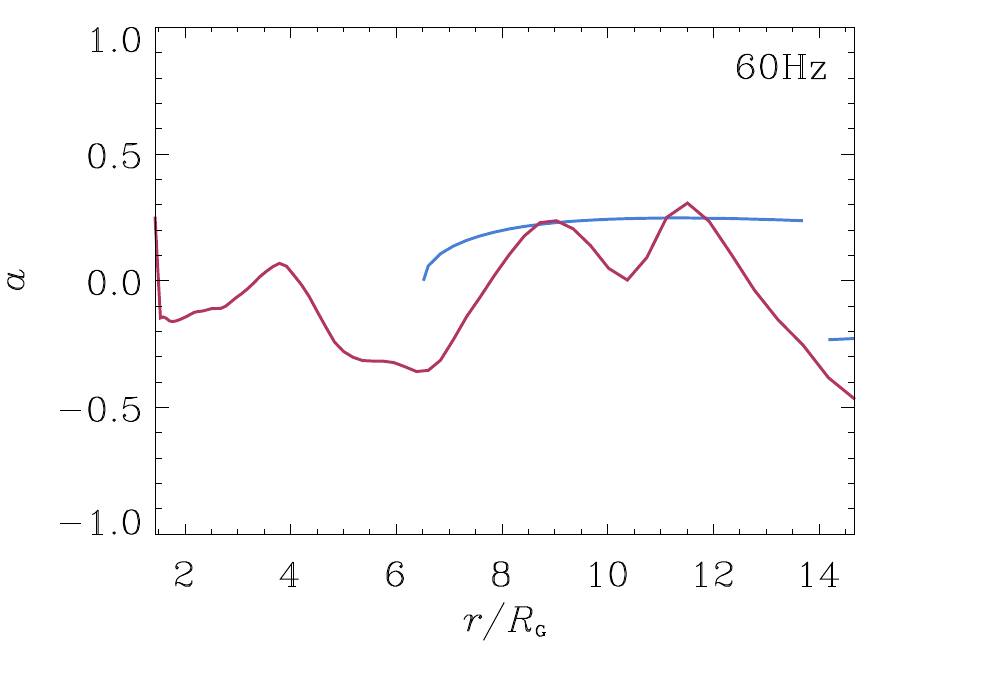}
\end{tabular*}
\caption[Local radial decay rate of prograde, $m=1$ power away from clump centers and diskoseismic
  predictions.]
 {Local radial decay rate $a$ (red; see Equation~\ref{eq:decay_rate}) in the prograde, $m=1$
  component of variability compared with the local radial decay constant $a_{\rm disp}$ (blue;
  positive inside corotation, negative outside corotation) predicted from the dispersion relation in
  Equation~\ref{eq:disp_rel} at specified frequencies in 90h (a), 910h (b), 915h (c and d),
  915h-64a (e), and 915h-64b (f).
  \label{fig:modedecay}}
\end{figure*}

Shock amplification and tilt aside, we characterize the nature of these apparently ubiquitous,
prograde spiral patterns by comparing the structure of their prograde components to diskoseismic
predictions. Isothermal diskoseismic models of untilted, flat disks indicate
that short wavelength WKB perturbations with
local radial wave number $k$ obey the dispersion relation \citep{oka87},
\begin{eqnarray}
k^2=\frac{(\hat{\omega}^2-\kappa_r^2)(\hat{\omega}^2-n\kappa_\theta^2)}{\hat{\omega}^2c_{\rm s}^2},
\label{eq:disp_rel}
\end{eqnarray}
where $\hat{\omega}\equiv\omega-m\Omega$ is the Doppler-shifted wave frequency, $\kappa_r$ is the 
radial epicyclic frequency, and $n$ is a non-negative integer that indicates the number of vertical 
nodes in the oscillation. Adopting the hypothesis that our prograde spiral patterns represent $n=0$, 
$m=1$ acoustic perturbations having angular frequency $\omega$, then the 
region around the corotation radius bounded by the inner and outer Lindblad resonances is an
evanescent region with local radial decay constant $a_{\rm disp}$ such that $k=ia_{\rm disp}$. If the
WKB approximation holds, we can directly calculate the local radial decay rate of the amplitude of
the prograde spiral using a derivative of the power in Figure~\ref{fig:mode_decomp}, specifically,
\begin{eqnarray}
a=\frac{1}{2}\frac{d}{dr}\left\{\log{\left[\frac{\omega}{2\pi}\frac{\langle A_\omega^2\rangle_\theta}{\langle\rho^2\rangle_{t,\theta,\phi}}\right]}\right\}.
\label{eq:decay_rate}
\end{eqnarray}
If our spirals are acoustic in nature, then up to its sign, this calculated radial decay rate should 
roughly match the theoretical decay constant $a_{\rm disp}$, that is, the imaginary component of the 
wave number from Equation~\ref{eq:disp_rel}. In Figure~\ref{fig:modedecay}, we plot both $a$ and 
$a_{\rm disp}$. While there is noise, it is noteworthy that the measured decay rate in each 
simulation is positive inside corotation, negative outside corotation, and approximately equal in 
magnitude to that predicted by the dispersion relation in Equation~\ref{eq:disp_rel}. We believe 
that this further supports the hypothesis that intraorbital power is the manifestation of spirals 
excited by orbiting clumps that are acoustic in nature. 

\begin{figure*}
\centering
\begin{tabular*}{\textwidth}{@{\extracolsep{\fill}}*{2}{@{}b{\figwd}}@{}}
\figimg{a}{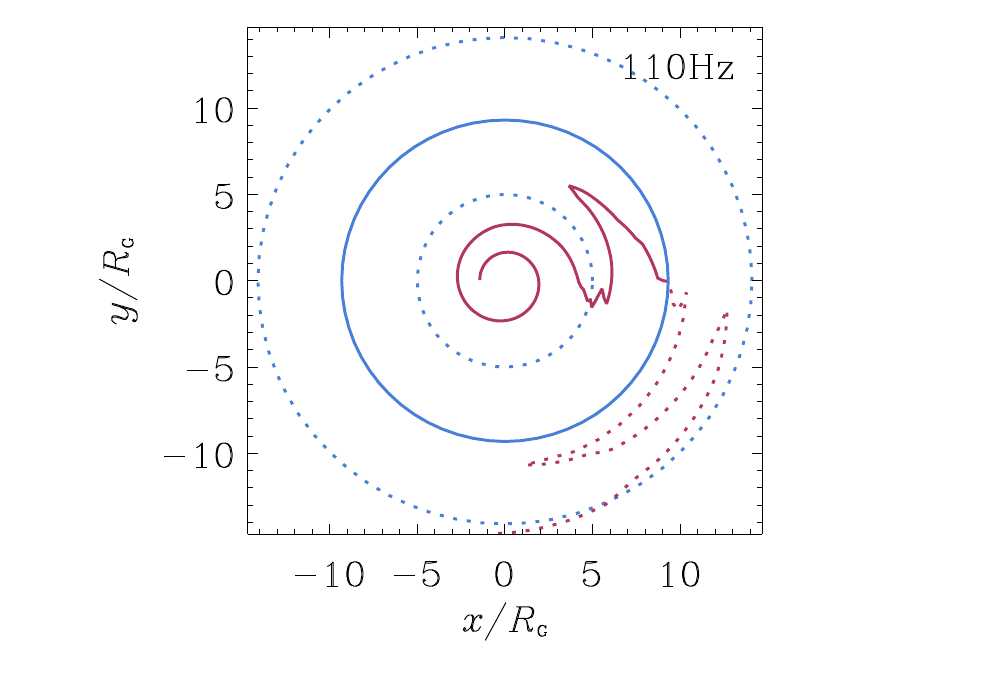} &
\figimg{b}{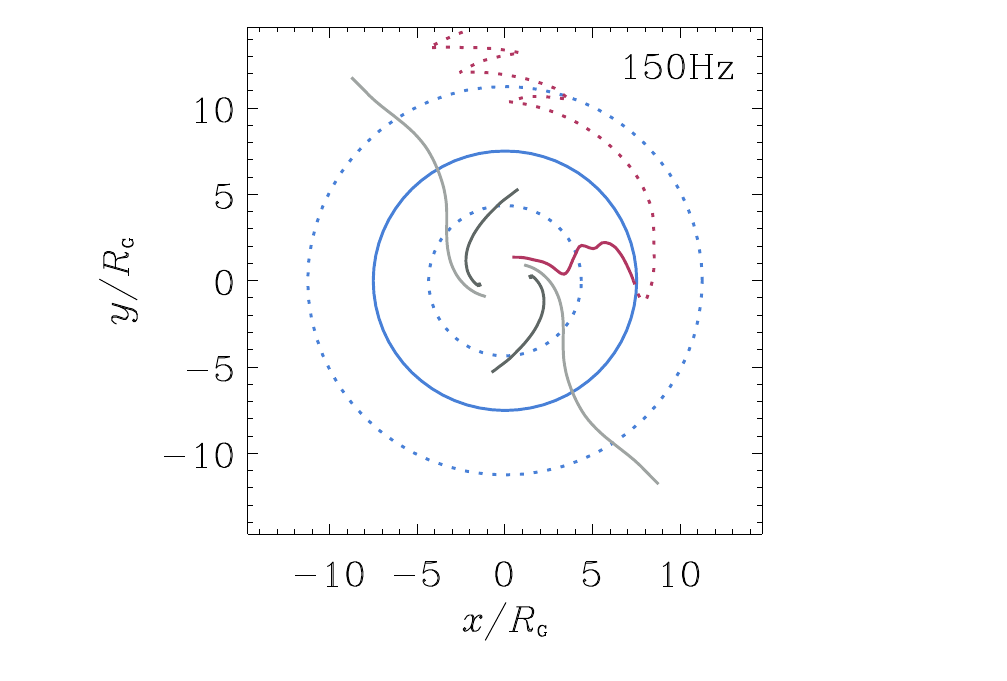} \\
\figimg{c}{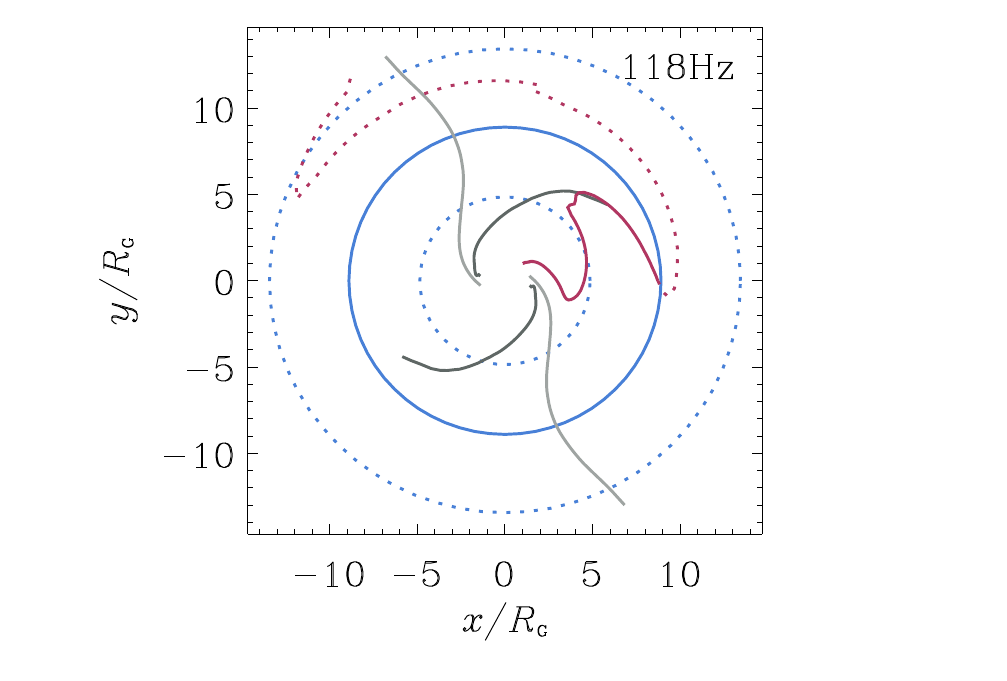} &
\figimg{d}{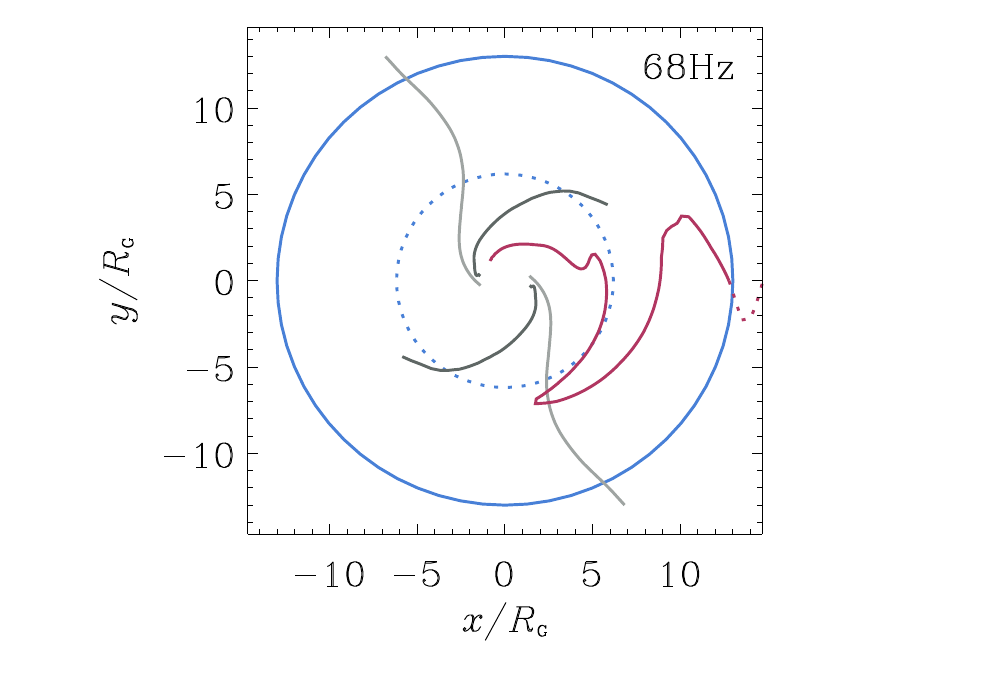} \\
\figimg{e}{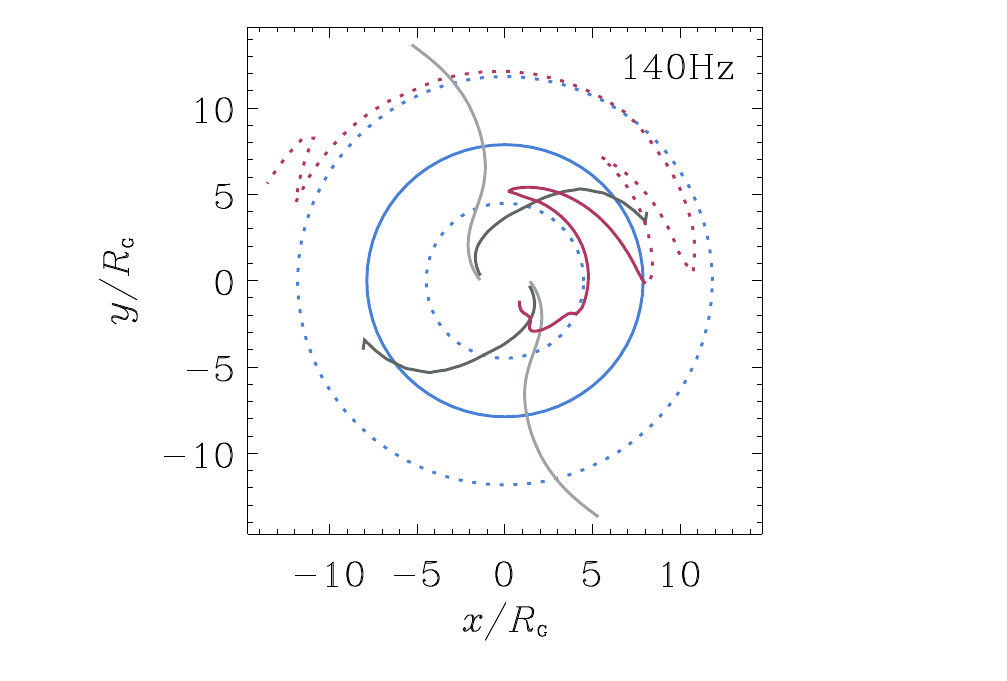} &
\figimg{f}{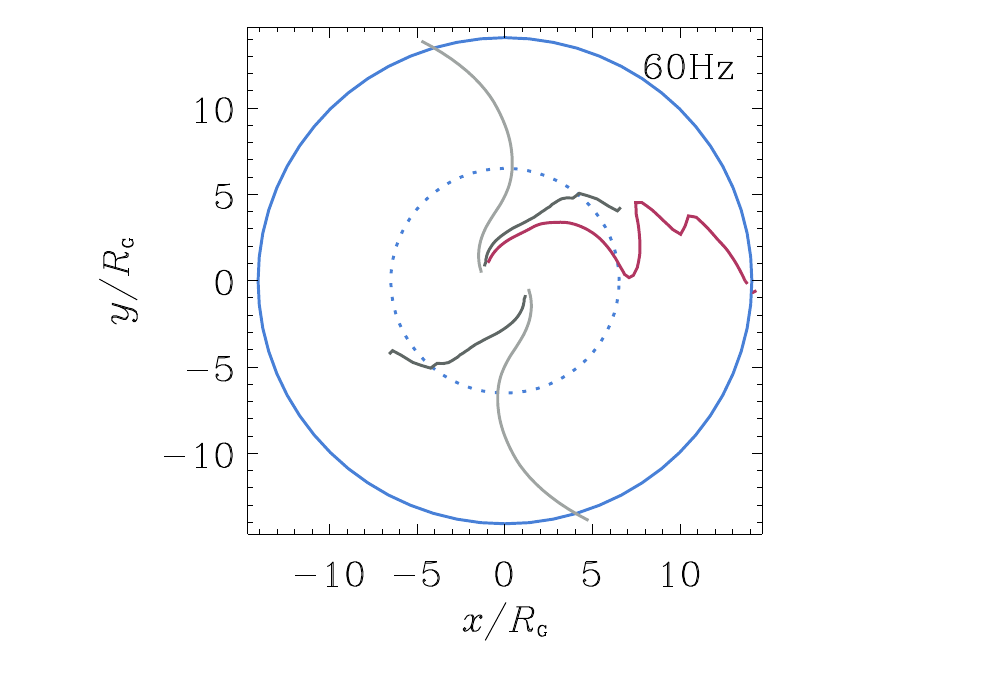}
\end{tabular*}
\caption[Snapshots of the spatial position of the maxima of the prograde component
  of $m=1$ mode power.]
 {Snapshots of the spatial position of the maxima of the prograde pattern of the $m=1$ mode power
  at each radius (red; solid inward and dotted outward of corotation) for patterns at the indicated
  frequencies in 90h (a), 910h (b), 915h (c and d),  915h-64a (e), and 915h-64b (f). For
  reference, we plot the approximate location of the background density maxima (light gray) and
  shock surfaces (dark gray) as well as circles marking the test-particle corotation radius (solid
  blue) and inner and outer Lindblad resonances (dashed blue). The prograde pattern rotates
  counterclockwise in time with respect to the density maxima and shock surfaces.
  \label{fig:mode_spiral}}
\end{figure*}

To aid in the visualization of these spiral patterns, Figure~\ref{fig:mode_spiral} depicts a
snapshot in time of the location (in an average sense) of the maxima of the prograde pattern at each
radius, that is,
$\overline{\alpha_\omega}\equiv\arctan{\left[\langle R_{\omega,\s}-I_{\omega,\c}\rangle_\theta/\langle R_{\omega,\c}+I_{\omega,\s}\rangle_\theta\right]}$
inspired by the definition of $\alpha_\omega$ in Equation~\ref{eq:pro_ret_funcs}. We interpret the 
shape of these curves as the average shape of the clump-spiral structure. One can imagine that these 
plots evolve in time by noting that the prograde spiral will rotate counterclockwise through the 
stationary background flow. Breaking down their structure, we see that the prograde spirals from 
each simulation share the same, qualitative, three-segment geometry: a trailing segment at small 
radii, a leading segment around an intermediate radius near the inner Lindblad resonance, and 
another trailing segment terminating at the corotating clump.

The first and last segments of these spirals seem consistent with the outward moving density 
fluctuations located in the over-dense arm and identified in Figure~\ref{fig:fourier3d_915h} by 
labels $2$, $3$, and $4$. Since $2$ and $3$ appear over roughly the same range of radii in their 
respective arms, our $m=1$ spiral model would predict that the two perturbations should be 
spatially symmetric though half a period out of phase in time. Looking closely however, we see that 
while $2$ is retreating away from the hole and $3$ is growing toward larger radii, they are both 
nonetheless simultaneously positive. This indicates that feature $2$ leads $3$ in phase by slightly 
less than half a period. This behavior must be due to $m>1$ contributions to the overall
clump/spiral structure, though we have not analyzed these higher order effects in detail.

Given the above, one might anticipate that the middle, leading segment ought to have an associated 
inward moving density enhancement in each over-dense arm. The absence of such features in 
Figure~\ref{fig:fourier3d_915h} has a two-fold explanation. First, Figure~\ref{fig:mode_spiral} only
roughly depicts the location of the spiral's maximum at each radius and not the maxima's relative
amplitudes. As seen in Figure~\ref{fig:mode_decomp}, the density 
perturbations ubiquitously reach local minima near the Lindblad resonances and, correspondingly, the 
leading segments of the spirals must be relatively weak. These perturbations therefore have little
effect in Figure~\ref{fig:fourier3d_915h}. Second, the retrograde component (not shown in 
Figure~\ref{fig:mode_spiral} for purposes of clarity) is oriented such that the whole leading 
segment reaches its maxima all at once and without any apparent motion. Although we strongly suspect 
that there is some connection between the common shape of these spirals and the location of the 
inner Lindblad resonances, it is unclear what mechanism might be responsible for the apparent 
correlations.

To summarize, orbiting clumps and their accompanying spiral patterns are generic in these 
simulations. Although these features do little to increase the overall variability in untilted disk 
geometries, the standing shocks characteristic of tilted simulations interact strongly with the 
orbiting spirals, producing regions of higher density where the two come into contact. Since the 
trailing segments of spiral pattern are more tightly wound than the standing shocks, the high 
density region appears to move outwardly along the shock surfaces. This mechanism explains the 
complex behavior shown in Figure~\ref{fig:fourier3d_915h} and provides an explanation for the 
characteristically higher variability at suborbital frequencies seen exclusively in tilted 
geometries. 

\section{Conclusions}
\label{sec:conclusions}

In our previous paper \citep{hen09}, we identified variability at frequencies comparable to those 
expected for inertial modes (i.e. below the local radial epicyclic frequency) in the inner parts of 
a GRMHD simulation of a tilted accretion flow (915h). This appeared to provide preliminary 
confirmation that warps and eccentricity might excite inertial modes even in the presence of MRI 
turbulence in accretion disks \citep{kats04,kat09,fer08}, whereas in shearing box and global 
untilted flow simulations, such inertial modes are destroyed by MRI turbulence \citep{arr06,rey09}. 
However, our more complete analysis here of simulation 915h, along with other tilted simulations 
910h and 915h-64, shows that the three-dimensional structure of the enhanced variability at 
suborbital frequencies is {\it not} consistent with the excitation of trapped inertial modes. 

The variability that we observe is in fact present in both untilted and tilted simulations, and 
originates from transient over-dense clumps of material co-orbiting with the background flow and 
randomly located at nearly all radii. Exhibiting a predominantly $m=1$ azimuthal structure, such 
fluctuations are likely only to manifest themselves in simulations which incorporate a complete 
$2\pi$ treatment of the azimuthal angle. We note that \cite{dol12} also identified $m=1$ structures 
in their global GRMHD simulations that appeared to be associated with the QPO they identified in 
their radiative post-processing of these simulations. It may be that these structures are similar to 
the clumps that we have identified in our work. In any case, the formation mechanism for these 
clumps remains unclear. Given that they orbit with the background flow, it may be that they are 
generated by some instability associated with a corotation resonance (e.g. \citealt{fu11}), but this 
requires further study.

We find that these orbiting clump structures are accompanied by extended acoustic spiral waves with 
pattern speeds equal to the orbital angular velocity of the clump. Exhibiting power which 
exponentially decreases away in radius from the orbit of the clump, these spirals are consistent 
with forced acoustic perturbations inside the otherwise evanescent region surrounding the corotation 
resonance \citep{oka87}. In the untilted simulation 90h, these spiral perturbations are quite weak. 
As shown by \cite{fra08}, however, converging eccentric orbits in tilted accretion flows produce two 
standing shock surfaces which precede two over-dense arms. The compression of the orbiting flow at 
these shocks amplifies the density amplitude of the otherwise weak acoustic spirals, thereby greatly 
enhancing their variability power signature. We therefore hypothesize that, given a greater tilt, 
any moderately thick accretion flow will exhibit greater variability at frequencies comparable to 
the orbital frequencies at all radii inward of perhaps $10$ to $15\RG$, that is, the furthest radial 
extent of the shocks. 

These findings are quite relevant to a recent class of models which produce a low-frequency QPO in a 
hot, somewhat tilted, inner flow that is qualitatively consistent with observations of black hole 
X-ray binaries \citep{ing09,ing11}. Such tilted flows naturally precess at a Lense-Thirring 
precession frequency determined by the hot flow's outermost edge \citep{fra07}. Therefore, in these 
low-frequency QPO models, we predict increased continuum variability at frequencies greater than the 
orbital frequency measured at either the hot flow's outer cutoff radius or the termination radius of 
the standing shocks, whichever is closer to the hole. 

Over short intervals of time, the variability due to the clumps and spirals occurs at discrete 
frequencies depending on which clumps happen to be present. While these frequencies 
are comparable to high-frequency QPOs in black hole X-ray binaries, the transient nature of the 
clumps implies that they cannot produce robust high-frequency QPOs, and are likely instead to just 
contribute to enhanced continuum variability as we noted above. We have not identified any other 
potential high-frequency QPO candidates in our simulations. We appear to be in the good company of 
many other accretion simulations to date, and we suspect that either the QPO is too weak to detect 
in current numerical simulations or, more likely, that the QPO requires physics which is not yet 
captured by our numerical simulations. As \cite{rey09} point out, current global simulations lack 
the necessary treatment of radiation physics to accurately simulate the near Eddington limited 
accretion which characterizes the steep power-law spectral state uniquely associated with the
high-frequency QPO. Therefore, in the coming years when new numerical methods meet more advanced 
computational technologies, variability studies akin to those we have presented here will become 
increasingly relevant in understanding the complex dynamics associated with black hole accretion 
flows. 

We thank Lars Bildsten for useful conversations, and Aleksey Generozov and Julia Wilson for help in 
analyzing the density compression ratios of the shocks. This work was supported in part by NSF grant 
AST-0707624. PCF acknowledges support from the National Science Foundation under grants AST-0807385
and PHY11-25915.

\bibliographystyle{apj}
\bibliography{references}

\end{document}